\documentclass{xjkas}
\def\year{2015} 
\def\month{October} 
\def\volume{00} 
\def\beginpage{1} 
\def\endpage{11} 
\def\received{July 27, 2015} 
\def\accepted{October 26, 2015} 
\date{Received \received; accepted \accepted}

\usepackage{flushend} 
\usepackage[breaklinks, colorlinks, citecolor=blue, linkcolor=blue, menucolor=blue, urlcolor=blue]{hyperref}

\title{
A Search for AGN Intra-day Variability with KVN\thanks{Part of a special issue on the Korean VLBI Network (KVN)}
}

\author[1]{Taeseok~Lee}
\author[1]{Sascha~Trippe}
\author[1]{Junghwan~Oh}
\author[2]{Do-Young~Byun}
\author[2]{Bong-Won~Sohn}
\author[2]{Sang-Sung~Lee}

\affil[1]{Department of Physics and Astronomy, Seoul National University, Gwanak-gu, Seoul 08826, Korea \email{tlee@astro.snu.ac.kr, trippe@astro.snu.ac.kr}}
\affil[2]{Korea Astronomy and Space Science Institute, Yuseong-gu, Daejeon 34055, Korea}

\def\corrauthor{S. Trippe}
\def\runningauthor{Lee et al.}

\def\runningtitle{A Search for AGN Intra-day Variability with KVN}

\def\keywords{galaxies: active --- galaxies: individual: 3C~111, 3C~279, 3C~454.3, BL Lac --- radiation mechanisms: non-thermal --- methods: statistical}

\def\abstracttext{
Active galactic nuclei (AGN) are known for irregular variability on all time scales, down to intra-day variability with relative variations of a few percent within minutes to hours. On such short timescales, unexplored territory, such as the possible existence of a shortest characteristic time scale of activity and the shape of the high frequency end of AGN power spectra, still exists. We present the results of AGN single-dish fast photometry performed with the Korean VLBI Network (KVN). Observations were done in a ``anti-correlated'' mode using two antennas, with always at least one antenna pointing at the target. This results in an effective time resolution of less than three minutes. We used all four KVN frequencies, 22, 43, 86, and 129 GHz, in order to trace spectral variability, if any. We were able to derive high-quality light curves for 3C~111, 3C~454.3, and BL Lacertae at 22 and 43 GHz, and for 3C~279 at 86 GHz, between May 2012 and April 2013. We performed a detailed statistical analysis in order to assess the levels of variability and the corresponding upper limits. We found upper limits on flux variability ranging from $\sim$1.6\% to $\sim$7.6\%. The upper limits on the derived brightness temperatures exceed the inverse Compton limit by three to six orders of magnitude. From our results, plus comparison with data obtained by the University of Michigan Radio Astronomy Observatory, we conclude that we have not detected source-intrinsic variability which would have to occur at sub-per cent levels.
}

\begin{document}
\jkashead 


\section{Introduction\label{sec:intro}}

Variability of the flux from active galactic nuclei (AGN) as function of time has been frequently reported at various time scales and wavelengths \citep[see, e.g.,][]{Trippe2011, Benlloch2001, Fuhrmann2008, Gupta2012}. Variability on a time scale under one day, so-called \emph{intra-day variability} (IDV), was first detected by \citet{Witzel1986}. AGN of various type, such as flat spectrum radio quasars (FSRQs), BL Lacertae (BL Lac) objects, and Seyfert galaxies have been observed to show IDV \citep{Lovell2008}.

\begin{table*}[t!]
\caption{Our target sources. \label{tab:source}}
\centering
\begin{tabular}{lcccc}
\toprule
 & 3C 111 & 3C 279 & 3C 454.3 & BL Lac \\
\midrule
R.A. (J2000) & 04:18:21.3 & 12:56:11.1  & 22:53:57.7 & 22:02:43.3\\
DEC (J2000)  & $+$38:01:36 & $-$05:47:22 & $+$16:08:54 & $+$42:16:40\\
Redshift     & 0.0485 & 0.537 & 0.859 & 0.0686 \\
Distance (Mpc) & 205 & 2996 & 5330 & 292\\
Type & Seyfert & QSO & Blazar & QSO\\
\bottomrule
\end{tabular}
\tabnote{Source data are from the NASA/IPAC Extragalactic Database (NED), adopting a cosmology with $H_0=73$\,km\,s$^{-1}$\,Mpc$^{-1}$, $\Omega_m$=0.27, ${\Omega}_{\Lambda}$=0.73.}

\end{table*}

\begin{table*}[t!]
\caption{Overview on observations in 2011 and 2012.\label{tab:obsstatus1}}
\centering
\begin{tabular}{llllcccc}
\toprule
Date & & & Source & Frequency [GHz] & Time (KY/KU/KT) [Hr] & Data Quality$^{\dagger}$ & $T_{\rm sys}$ [K] \\
\midrule

2011 & Dec & 30& 3C 279$^{\rm *}$ & 22 & ---/---/5.9 & ---/---/high & ---/---/110\\
	 &     &   &    & 43 & ---/---/5.9 & ---/---/high & ---/---/180\\
	 &     &   &    & 86 & 5.9/5.8/--- & high/high/--- & 230/260/--- \\
	 &     &   &    & 129& 5.9/5.8/--- & high/high/--- & 230/400/--- \\	 \addlinespace
	 
	&     &&BL Lac & 22  & 2.7/2.4/0.9 & low/low/high & 70/80/110 \\ 
	&     &&       & 43  & 2.7/2.4/0.9 & low/low/high & 160/120/180\\ 
	&     &&       & 86  & 2.7/2.4/--- & low/low/--- & 170/240/--- \\ 
	&     &&       & 129 & 2.7/2.4/--- & low/low/--- & 160/350/---\\ \addlinespace
 \addlinespace
2012 & Jan &26& 3C 279 & 22 & ---/---/7.4 & ---/---/high & ---/---/120 \\
     &     &  &     & 43 & ---/---/7.4 & ---/---/high & ---/---/180 \\
	 &	   &&	   & 86 & 7.4/7.1/--- & low/low/--- & 190/210/--- \\
	 &     & &      & 129& 7.4/7.1/--- & low/low/--- & 410/590/--- \\  \addlinespace
	 
	 &     & &BL Lac& 22 & ---/---/7.1 & ---/---/high & ---/---/100 \\
     &     & &      & 43 & ---/---/7.1 & ---/---/high & ---/---/140 \\
     &     & &      & 86 & 6.7/3.8/--- & low/low/--- & 170/180/--- \\
	 &     & &      & 129& 6.7/3.8/--- & low/low/--- & 160/200/--- \\  \addlinespace

     & Feb & 21&3C 279 & 86 & 7.9/7.9/--- & high/low/--- & 230/400/--- \\
	 &     &   &    & 129& 7.9/7.9/--- & high/low/--- & 300/760/--- \\  \addlinespace

     &     & &BL Lac& 86 & 2.9/2.9/--- & low/low/--- & 150/150/--- \\
	 &     & &      & 129& 2.9/2.9/--- & low/low/--- & 290/580/--- \\  \addlinespace

	 & Apr & 30&3C 279 & 22 & ---/---/4.6 & ---/---/low & ---/---/190 \\
     &     &   &    & 43 & ---/---/4.6 & ---/---/low & ---/---/190 \\
     &     &   &    & 86 & 6.9/---/--- & high/---/--- & 520/---/--- \\
	 &     &   &    & 129& 6.9/---/--- & low/---/---  & 1200/---/--- \\  \addlinespace

     &     & &BL Lac& 86 & 3.1/---/--- & low/---/--- & 190/---/--- \\
     &     &  &     & 129& 3.1/---/--- & low/---/--- & 190/---/--- \\ \addlinespace

     & May & 31&3C 279 & 22 & ---/---/6.9 & ---/---/high & ---/---/240 \\
     &     &   &    & 43 & ---/---/6.9 & ---/---/high & ---/---/170 \\
     &     &   &   & 86 & 6.8/6.9/--- & high/high/---& 460/470/--- \\
	 &     &   &    & 129& 6.8/6.9/--- & low/low/---  & 1100/1300/--- \\  \addlinespace

     &     & &BL Lac& 22 & ---/---/3.9 & ---/---/high & ---/---/130 \\
     &     & &      & 43 & ---/---/3.9 & ---/---/high & ---/---/250 \\
     &     & &      & 86 & 3.4/3.8/--- & low/low/---  & 280/290/--- \\
	 &     & &      & 129& 3.4/3.8/--- & low/low/---  & 730/600/--- \\ \addlinespace

     & Nov & 28/29/30&3C 111 & 22 & 27/---/25 & high/---/high & 70/---/90 \\
     &     & &      & 43 & 27/---/25 & low/---/high & 140/---/120 \\  \addlinespace

     &     & &3C 279$^{\rm *}$ & 22 & 16/---/17 & high/---/high & 70/---/140 \\
     &     & &      & 43 & 16/---/17 & high/---/high & 150/---/200 \\ 
\bottomrule
\end{tabular}
\tabnote{ 
$^{\dagger}$Referring to the agreement between cross-scan source profile and theoretical Gaussian \\ 
$^{\rm *}$Data suffer from systematic errors due to technical failures (cf. Section \ref{sec:obs})
}
\end{table*}

\begin{table*}[t!]
\caption{Overview on observations in 2013.\label{tab:obsstatus2}}
\centering
\begin{tabular}{llllcccc}
\toprule
Date & && Source & Frequency [GHz] & Time (KY/KU/KT) [Hr] & Data Quality$^{\dagger}$ & $T_{\rm sys}$ [K]\\
\midrule 
2013 & Feb & 22/23/24&3C 111 & 22 & 29/---/20 & high/---/high & 70/---/80 \\
     &     &       && 43 & 29/29/20  & high/low/high & 130/200/120 \\
     &     &       && 86 & ---/29/--- & ---/low/---  & ---/200/--- \\ \addlinespace
     
     & Apr &18&3C 454.3& 22 & 3.8/4.9/4.0 & high/high/high & 100/100/150 \\
     &     &  &     & 43 & 3.8/---/4.0 & high/---/high & 150/---/150 \\
     &     &  &     & 86 & ---/4.9/--- & ---/low/---  & ---/260/--- \\ \addlinespace
	 
     &     & &BL Lac& 22 & 5.6/2.9/5.6 & high/high/high & 70/120/90 \\
     &     & &      & 43 & 5.6/2.9/5.6 & high/low/high & 190/110/120\\ \addlinespace
 
     & May & 18&3C 111& 22 & ---/---/9.0 & ---/---/high & ---/---/160 \\
     &     &   &    & 43 & ---/---/9.0 & ---/---/high & ---/---/140 \\
     &     &   &    & 86 & 9.6/8.7/--- & low/low/---  & 210/210/--- \\
	 &     &   &    & 129& 9.6/8.7/--- & low/low/---  & 980/260/--- \\  \addlinespace

     &     &&3C 454.3& 22 & ---/---/3.9 & ---/---/low & ---/---/160 \\
     &     &&       & 43 & ---/---/3.9 & ---/---/low & ---/---/140 \\
     &     &&       & 86 & 3.9/3.9/--- & low/low/---  & 110/180/--- \\
	 &     &&       & 129& 3.9/3.9/--- & low/low/---  & 1000/200/--- \\ \addlinespace

     & Oct &17&4C 69.21& 22 & 5.2/---/--- & high/---/--- & 100/---/--- \\
     &     &  &     & 43 & 5.2/---/--- & low/---/--- & 200/---/--- \\
     &     &  &     & 86 & ---/9.8/9.8 & ---/low/low  & ---/270/230 \\
	 &     &  &     & 129& ---/9.8/9.8 & ---/low/low  & ---/330/300 \\  \addlinespace
 
     &     & &BL Lac& 86 & ---/6.3/7.7 & ---/low/low  & ---/260/320 \\
	 &     & &      & 129& ---/6.3/7.7 & ---/low/low  & ---/290/500 \\ \addlinespace

     & Nov & 1&3C 454.3& 86 & 6.6/---/6.3 & low/---/low  & 260/---/280 \\
	 &     &  &      & 129& 6.6/---/6.3 & low/---/low  & 650/---/480 \\  \addlinespace
	 
     &     & &4C 69.21& 86 & 8.8/---/8.8 & low/---/low  & 200/---/430 \\
	 &     & &       & 129& 8.8/---/8.8 & low/---/low  & 390/---/860 \\ \addlinespace
 
     & Dec & 21&3C 111 & 22 & ---/---/3.8 & ---/---/high & ---/---/70 \\
     &     &   &    & 43 & ---/---/3.8 & ---/---/high & ---/---/110 \\
     &     &   &    & 86 & 3.8/3.8/--- & high/low/---  & 150/160/--- \\
	 &     &   &    & 129& 3.8/3.8/--- & low/low/---  & 130/150/--- \\  \addlinespace
	 
     &     & &3C 454.3$^{\rm *}$& 22 & ---/---/5.1 & ---/---/high & ---/---/80 \\
     &     & &      & 43 & ---/---/5.1 & ---/---/high & ---/---/110 \\
     &     & &      & 86 & 5.1/5.1/--- & high/high/---  & 150/180/--- \\
	 &     & &      & 129& 5.1/5.1/--- & low/low/---  & 140/170/--- \\  \addlinespace

	 &     & &4C 69.21 & 22 & ---/---/1.9 & ---/---/high & ---/---/90 \\
     &     & &        & 43 & ---/---/1.9 & ---/---/low  & ---/---/150 \\
     &     & &        & 86 & 1.9/1.9/--- & low/low/---  & 190/200/--- \\
	 &     & &        & 129& 1.9/1.9/--- & low/low/---  & 180/250/--- \\ 
\bottomrule
\end{tabular}
\tabnote{
$^{\dagger}$Referring to the agreement between cross-scan source profile and theoretical Gaussian \\ 
$^{\rm *}$Data suffer from systematic errors due to technical failures (cf. Section \ref{sec:obs})
}
\end{table*}


\begin{table*}[t!]
\caption{The variability parameters.\label{tab:param}}
\centering
\begin{tabular}{lllccccccc}
\toprule
Date & & Source & Frequency$^{\rm a}$ & Time$^{\rm b}$& m & $\chi^2_r$ & N$^{\rm c}$ &$\chi^2_{r(<0.1\%)}$$^{\rm d}$& $T_b$ \\
\midrule \addlinespace
2012 & May & 3C 279 & 86 & 0.28 & 4.8 & 0.8 & 208 &1.3& $<2.1\times 10^{17} $ \\ \addlinespace
2012 & Nov & 3C 111 & 22 & 0.33 & 4.1 & 1.2 & 220 &1.3& $<3.5\times 10^{15} $ \\
     &     &       &    & 0.41 & 3.7 &  1.2& 286 &1.3& $<1.9\times 10^{15} $ \\
     &     &       &    & 0.41 & 4.3 & 1.4& 185 &1.4& $<2.1\times 10^{15} $ \\  \addlinespace
2013 & Feb & 3C 111 & 22 & 0.37 & 2.8 & 1.1& 128 &1.4& $<2.7\times 10^{15} $ \\
     &     &       &    & 0.41 & 3.0 & 1.3& 242 &1.3& $<2.3\times 10^{15} $ \\
     &     &       &    & 0.41 & 3.4 & 1.1& 211 &1.3& $<2.8\times 10^{15} $ \\
     &     &       & 43 & 0.37 & 6.1 & 1.7 & 120 &1.4& $<1.5\times 10^{15} $  \\
     &     &       &    & 0.41 & 6.8 & 1.2& 130 &1.4& $<1.2\times 10^{15} $  \\
     &     &       &    & 0.32 & 6.5 & 1.4 & 97 &1.5& $<1.8\times 10^{15} $  \\ \addlinespace
2013 & Apr & 3C 454.3 & 22 & 0.16 & 5.1 & 0.7 & 96 &1.5& $<2.9\times 10^{18} $  \\
     &     &         &    43 & 0.16 & 7.6 & 1.5 & 94 &1.5& $<1.1\times 10^{18} $ \\
 	 &     & BL Lac  & 22    & 0.23 & 1.6 & 0.6 & 82 &1.6& $<1.8\times 10^{16} $  \\
     &     &         &    43 & 0.23 & 2.5 & 0.9 & 83 &1.5& $<7.2\times 10^{15} $ \\

\bottomrule
\end{tabular}
\tabnote{
$^{\rm a}$in units of GHz; 
$^{\rm b}$in units of days; 
$^{\rm c}$number of data points;
$^{\rm d}$reduced $\chi^2$ for a signal with a false alarm probability of 0.1\%
}
\vskip1.5em
\end{table*}

Theoretical models constructed for the explanation of IDV mechanisms include accretion disk models with flares or disturbances \citep{Abramowicz1991}, Doppler boosted relativistic jets \citep{Blandford_Konigl1979}, and interstellar scintillation (ISS) \citep{Rickett1990}. On the one hand, interstellar scintillation  is a source-extrinsic mechanism, due to ionized interstellar medium (de)focusing the light traveling from an AGN to Earth. \citet{Lovell2008} claimed that at 5 GHz the known short term variability can be caused by scintillation. However, ISS becomes irrelevant factor at higher frequency radio bands as it scales with the observation frequency, $\nu$, like $\nu^{-2.2}$. Hence, ISS should be unimportant at $>$22 GHz. On the other hand, observed brightness temperatures well above the inverse Compton limit suggest Doppler boosting accompanied by rapid intrinsic variability \citep{Kellermann_Pauliny-Toth1969}. 

On the shortest time scales, down to a few minutes, the variability of AGN is only poorly probed, implying the need for further investigation for various reasons.
Firstly to probe the possible existence of a shortest time scale of AGN activity, possibly given by shocks in relativistic jets \citep{Blandford_Konigl1979, Marscher_Gear1985}. Secondly, the AGN power spectrum at the highest sampling frequency needs to be probed. Though, on time scale between hours and years, those power spectra are known to obey red noise statistics \citep{Park2014}, there is still unknown territory at shorter time scales.
In addition, the degree of simultaneity of flux variations across frequency bands ought to be studied. Even though a characteristic time delay between lightcurves obtained at different frequencies due to the variations of the optical depth with frequency is expected by certain emission model based on expanding AGN outflows, this is only found sometimes and on few-hour scales \citep{Marscher_Gear1985}.

For spectroscopic observations with time resolutions of minutes, the Korean VLBI Network (KVN) is the tool of choice. KVN has three identical 21-meter antennas located at three different sites, each equipped with four receivers which can operate simultaneously at up to four frequencies: 22, 43, 86, and 129 GHz.
In this paper, we investigate the variability on time scales of few minutes of four AGN: 3C~111, 3C~279, 3C~454.3, and BL Lacertae. Several observing runs were conducted from 2011 to 2013. We derived upper limits on source variability and the corresponding upper limits on brightness temperature.

\section{Observations\label{sec:obs}}

We performed observations with the KVN 21-meter radio telescopes at Yonsei, Ulsan, and Tamna from December 2011 to December 2013. We used all three antennas as independently operating single dishes at all four frequencies of 22, 43, 86, and 129 GHz, partially simultaneously. Jupiter, Venus, and Mars were chosen as amplitude calibrators, since they are stable radio emission sources. We pointed each antenna onto the target for five minutes, then five minutes onto the calibrator. We used a cross scan observation mode to obtain the flux density of the sources. One scan sequence consists of four scans, i.e., back and forth scans in azimuthal direction and in elevation direction, respectively. 

To examine the variability on time scale on the order of minutes, it is essential to obtain densely sampled uninterrupted light curves. In order to achieve this, we paired antennas with their receivers tuned to the same frequency and polarization. Whenever one antenna needed to point at the calibrator, the other one was pointed at the target. The maximum slew speed of KVN is three degree per second and the maximum acceleration is three degree per square second, which also helped us to minimize the inevitable time consumption for switching between target and calibrator.

\begin{figure*}[t!]
\centering
\includegraphics[width=80mm]{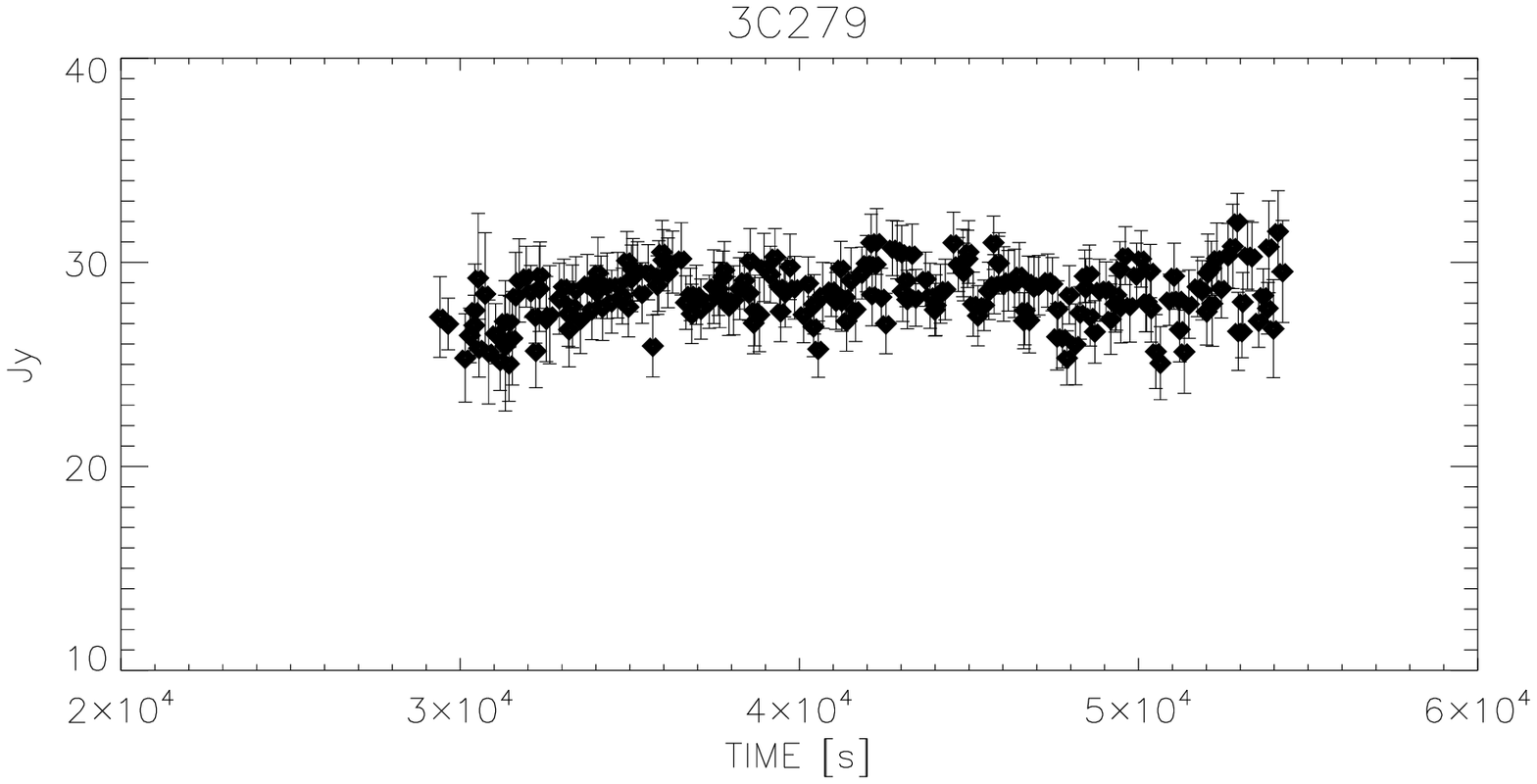} \hskip5mm 
\includegraphics[width=80mm]{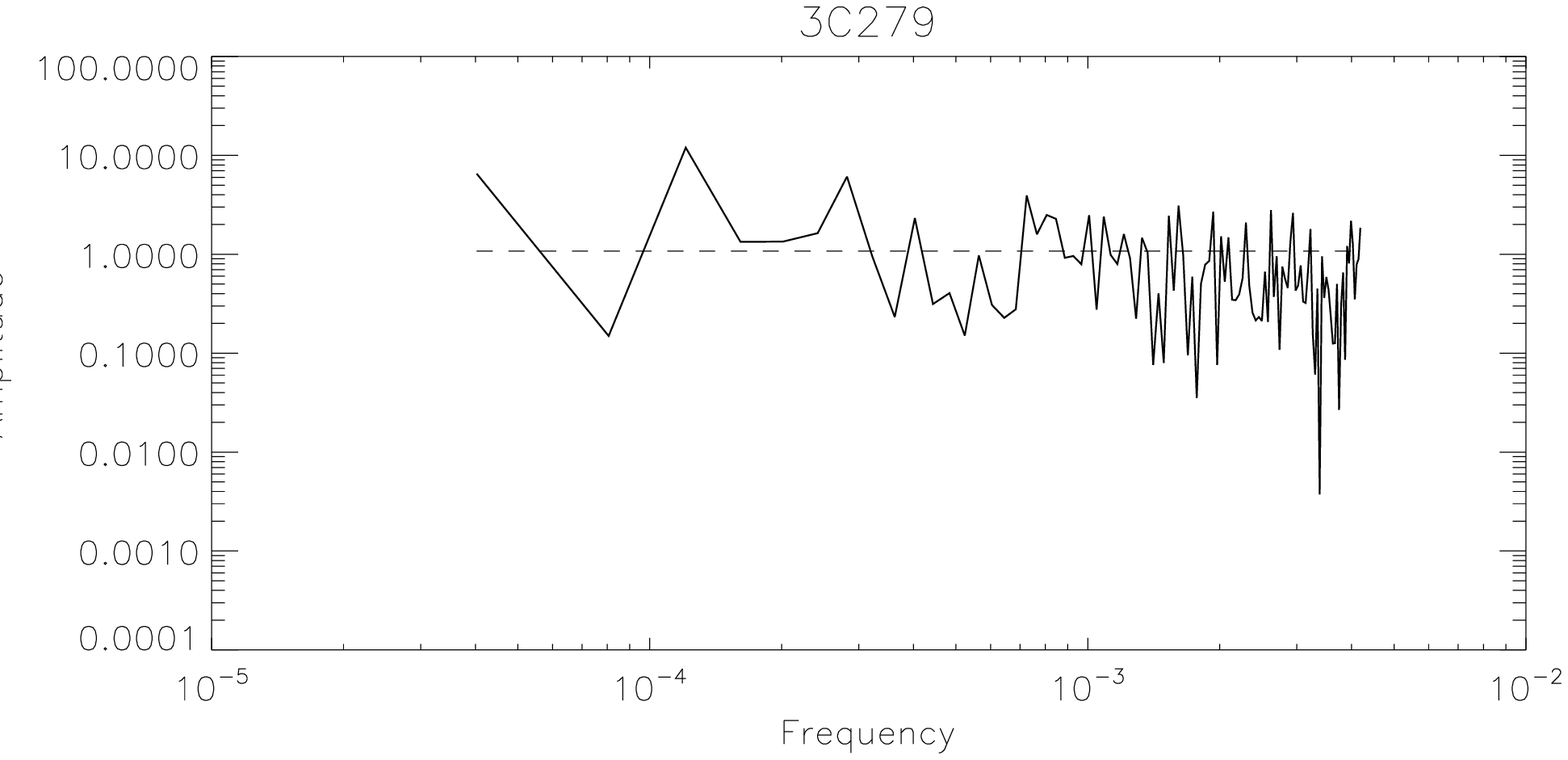}
\caption{The 86-GHz lightcurve (left) and power spectrum (right) of 3C279 as observed in May 2012. The dashed line in the power spectrum indicates the mean value; the sampling frequency is in units of s$^{-1}$. Note that the axis scales are logarithmic, leading to downward spikes being more pronounced that upward deviations.  \label{fig:3c279}}
\end{figure*}

\begin{figure*}[t!]
\centering
\includegraphics[width=80mm]{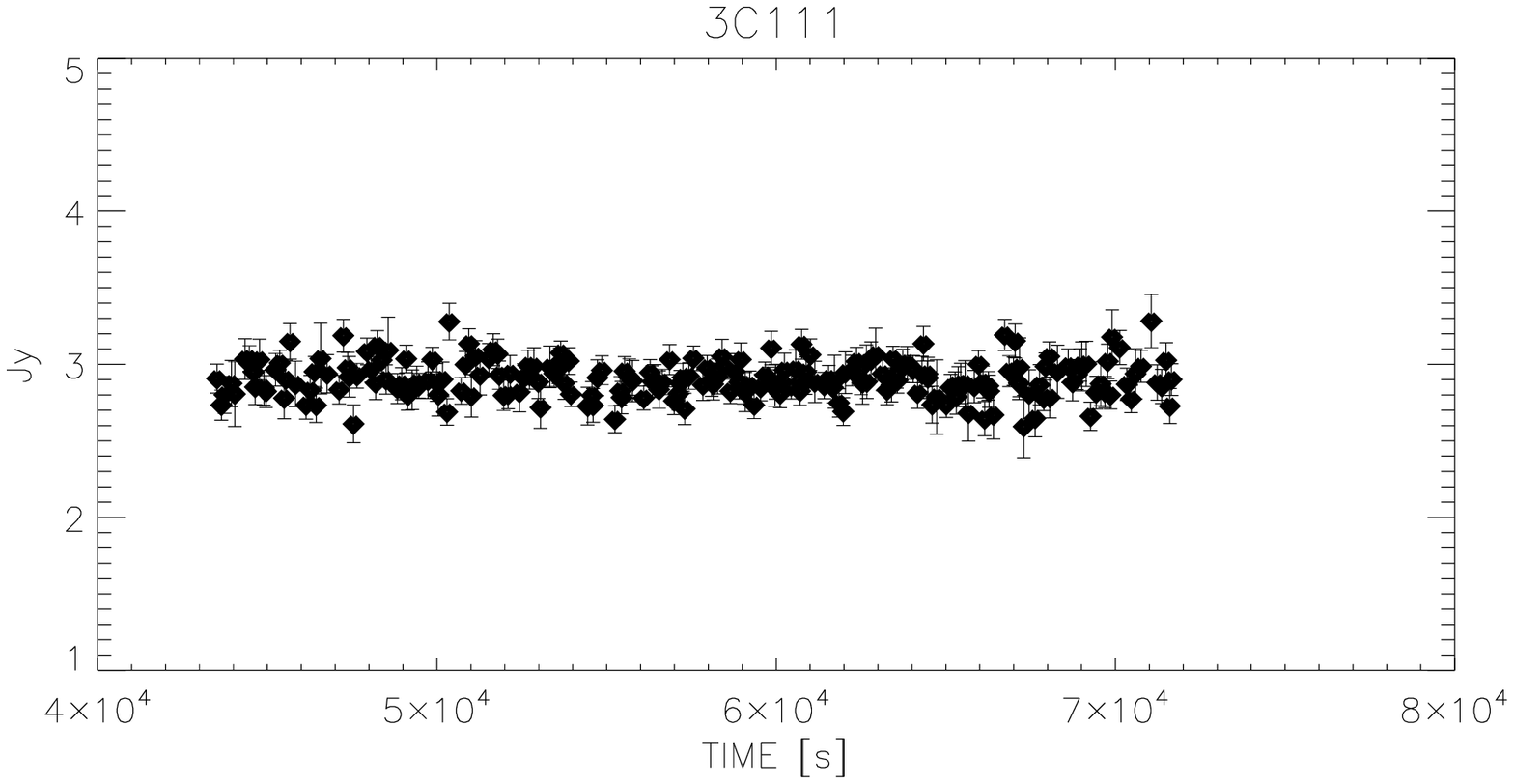} \hskip5mm 
\includegraphics[width=80mm]{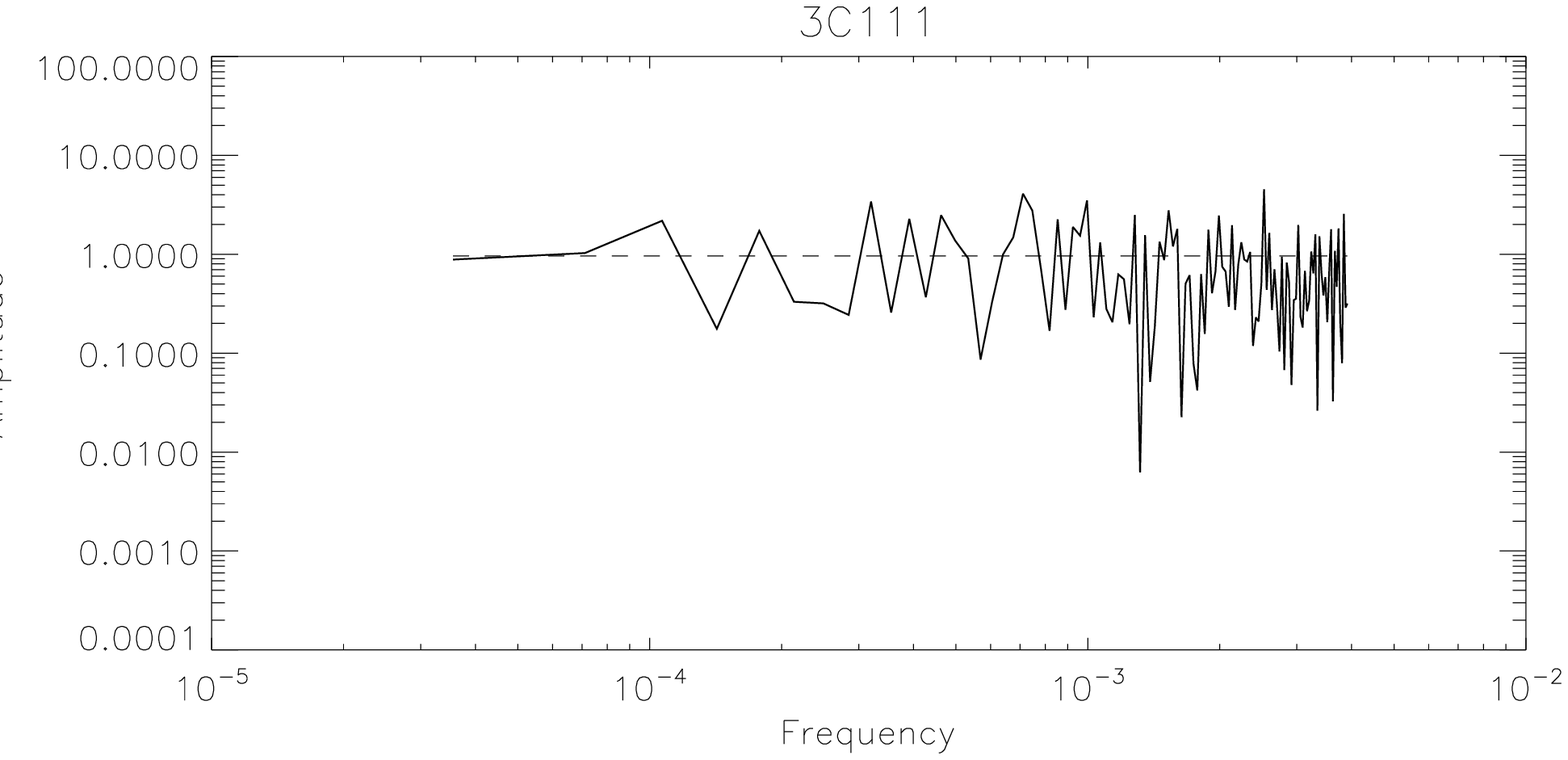} \\  
\includegraphics[width=80mm]{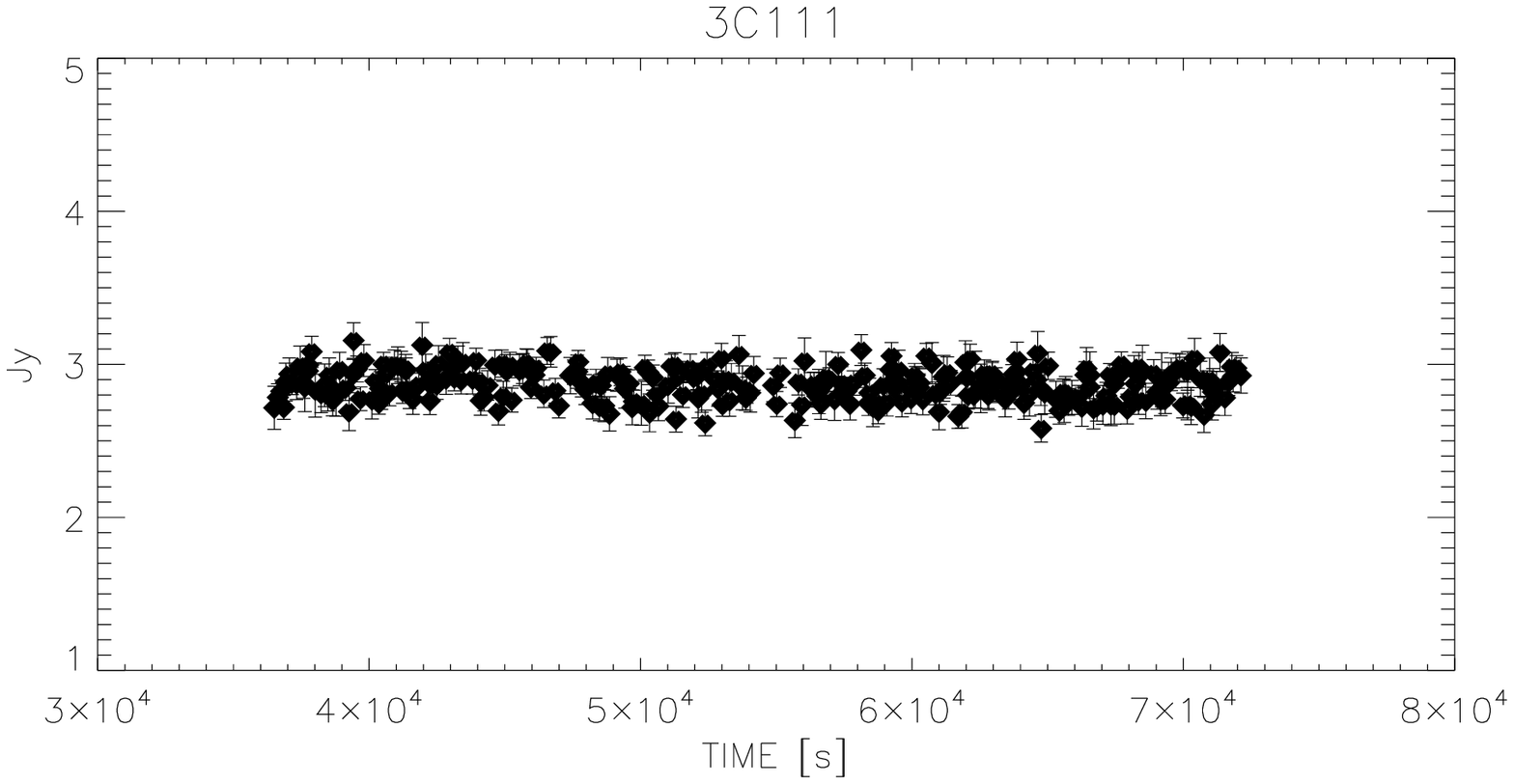} \hskip5mm 
\includegraphics[width=80mm]{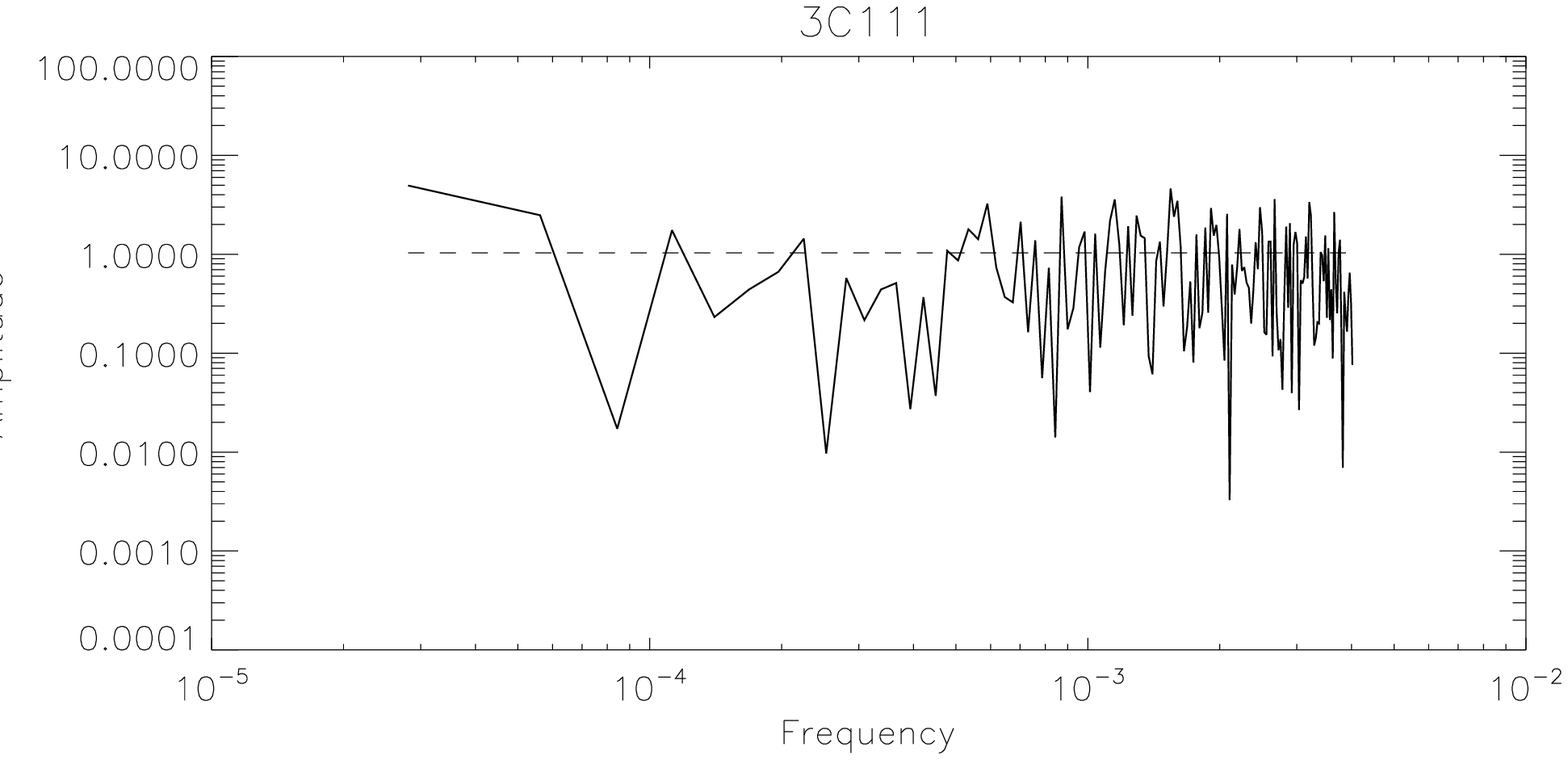} \\  
\includegraphics[width=80mm]{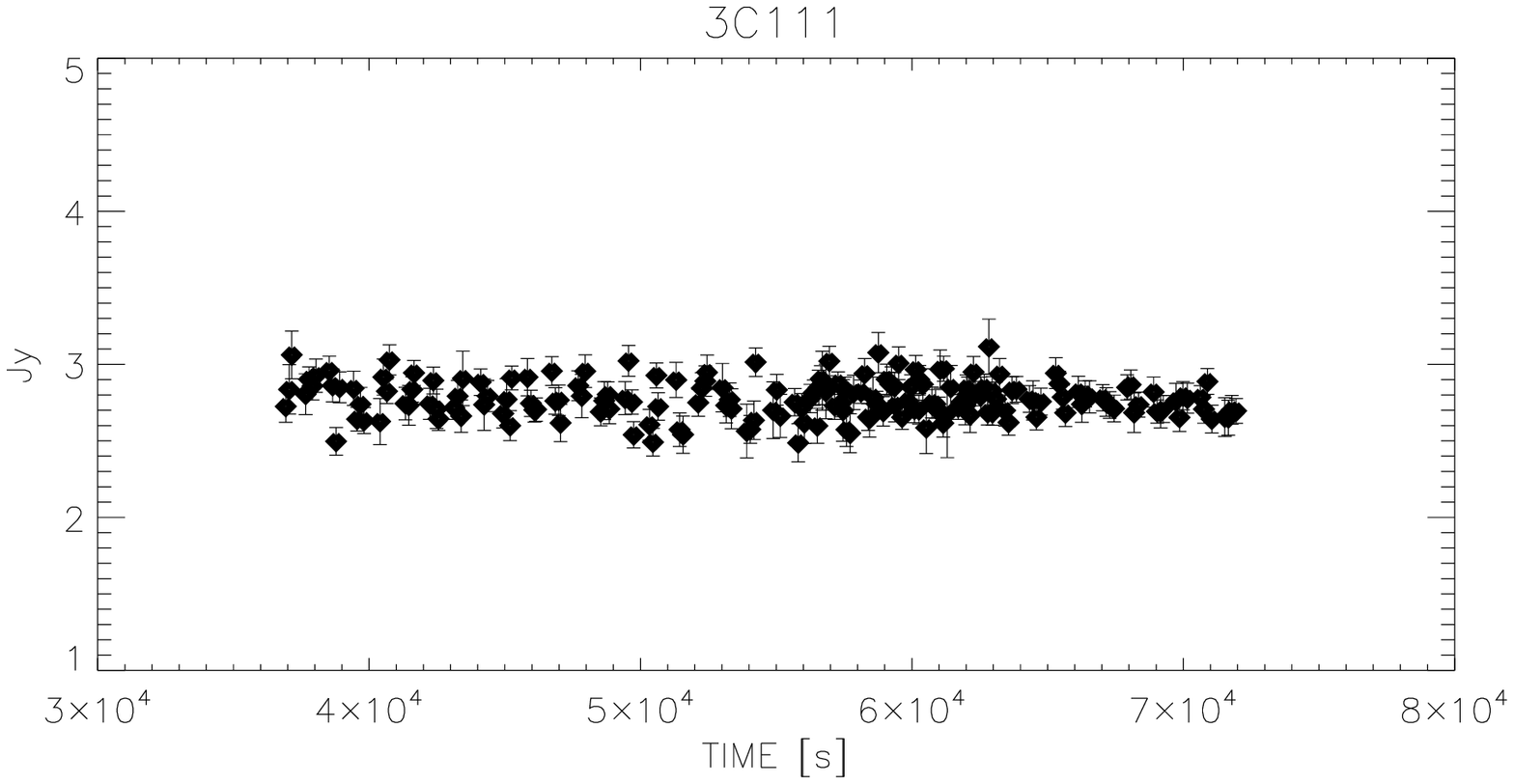} \hskip5mm 
\includegraphics[width=80mm]{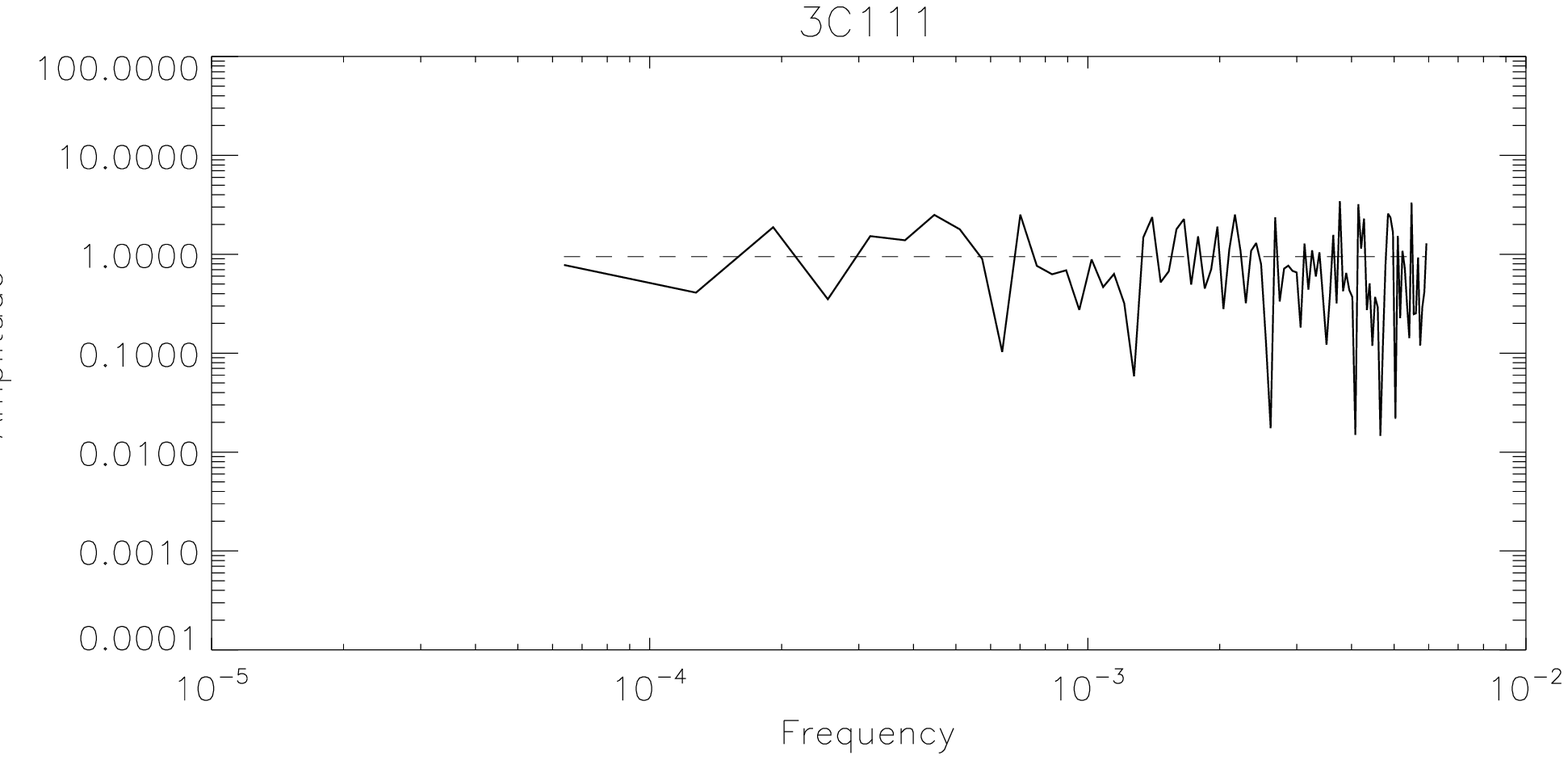}
\caption{The 22-GHz lightcurves and power spectra of 3C~111 observed during three consecutive days in November 2012. Dashed lines in the power spectra indicate mean values; the sampling frequency is in units of s$^{-1}$. \label{fig:3c111Nov}}
\end{figure*}

\begin{figure*}[t!]
\centering
\includegraphics[width=80mm]{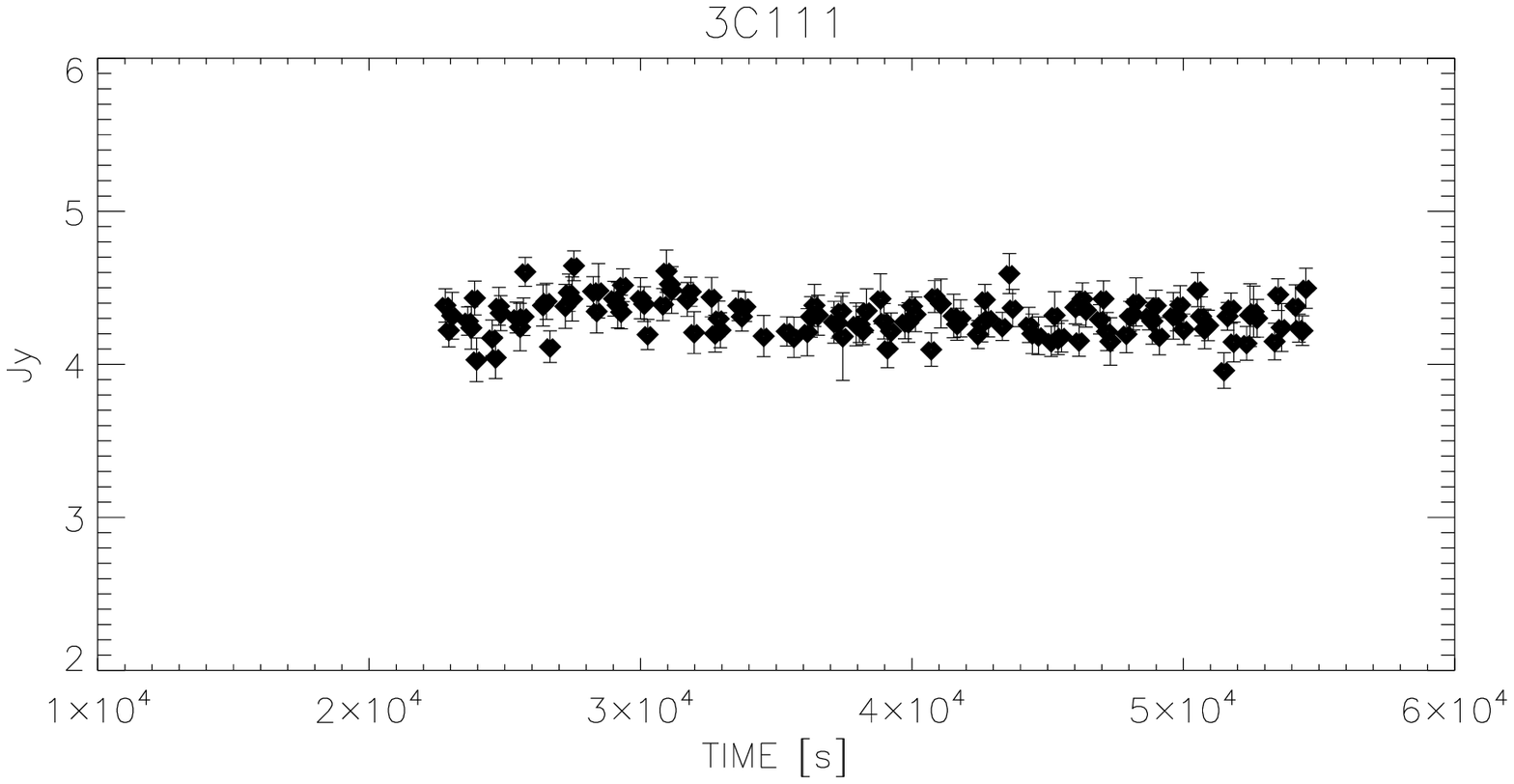} \hskip5mm 
\includegraphics[width=80mm]{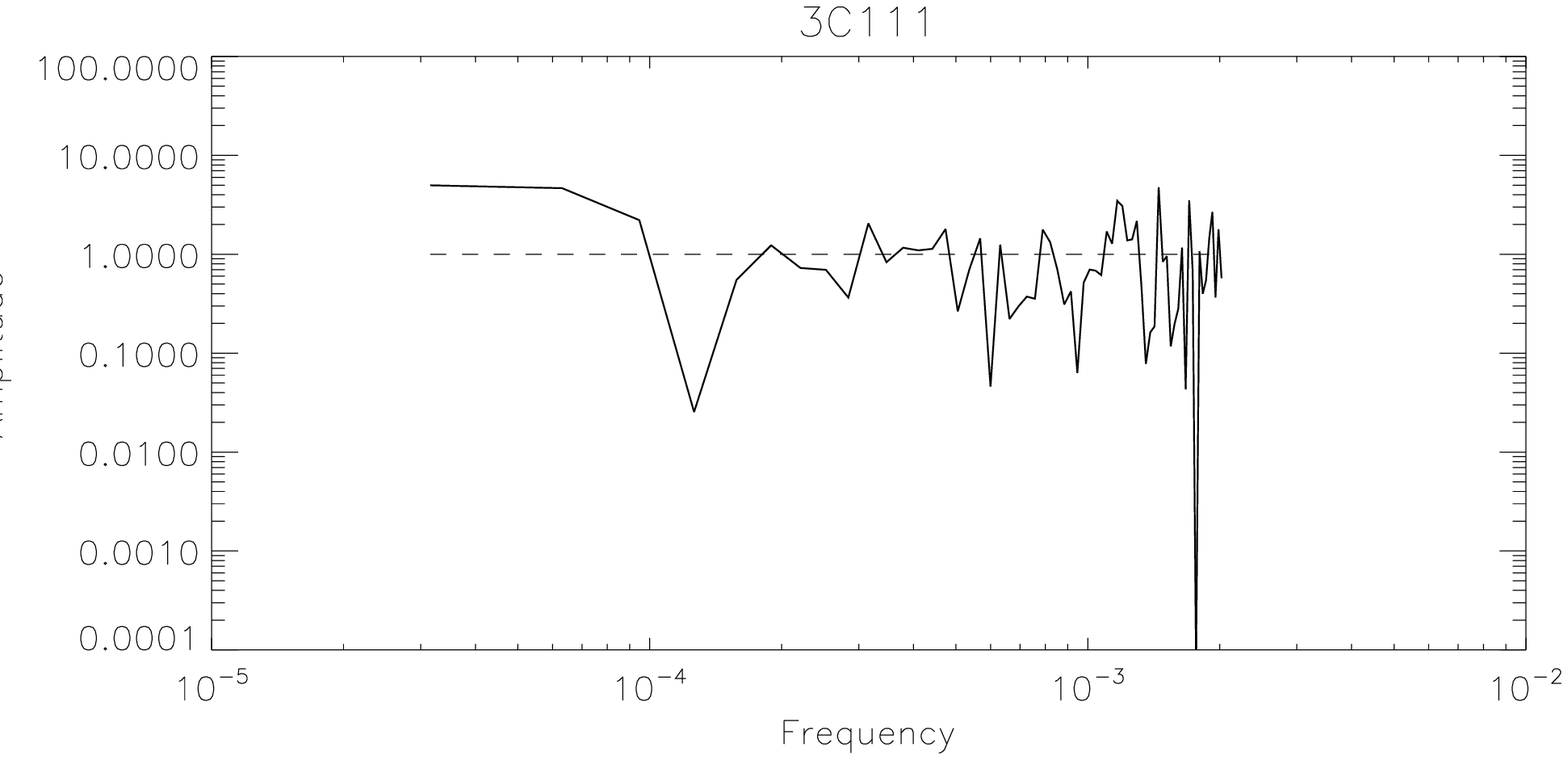} \\  
\includegraphics[width=80mm]{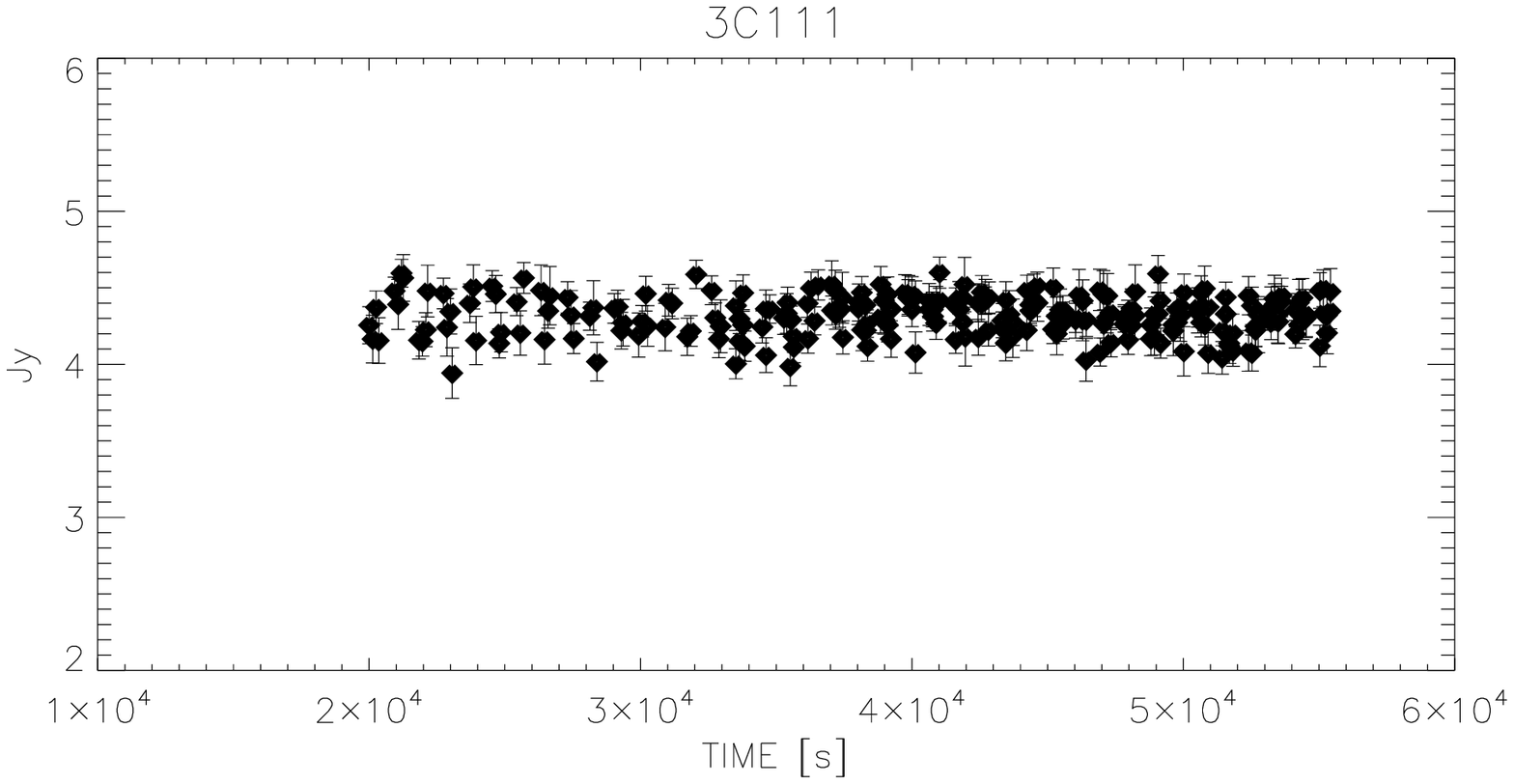} \hskip5mm 
\includegraphics[width=80mm]{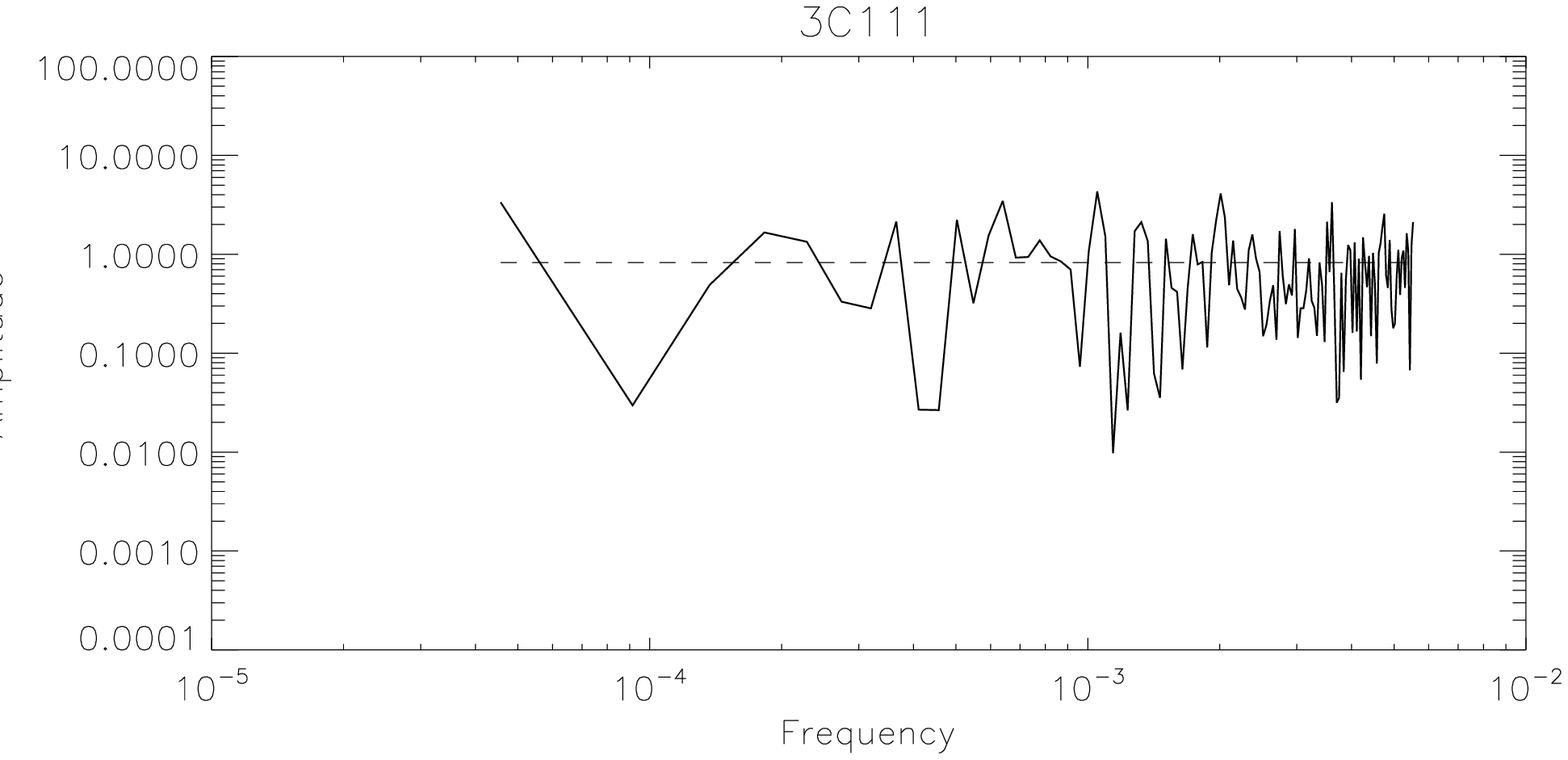} \\  
\includegraphics[width=80mm]{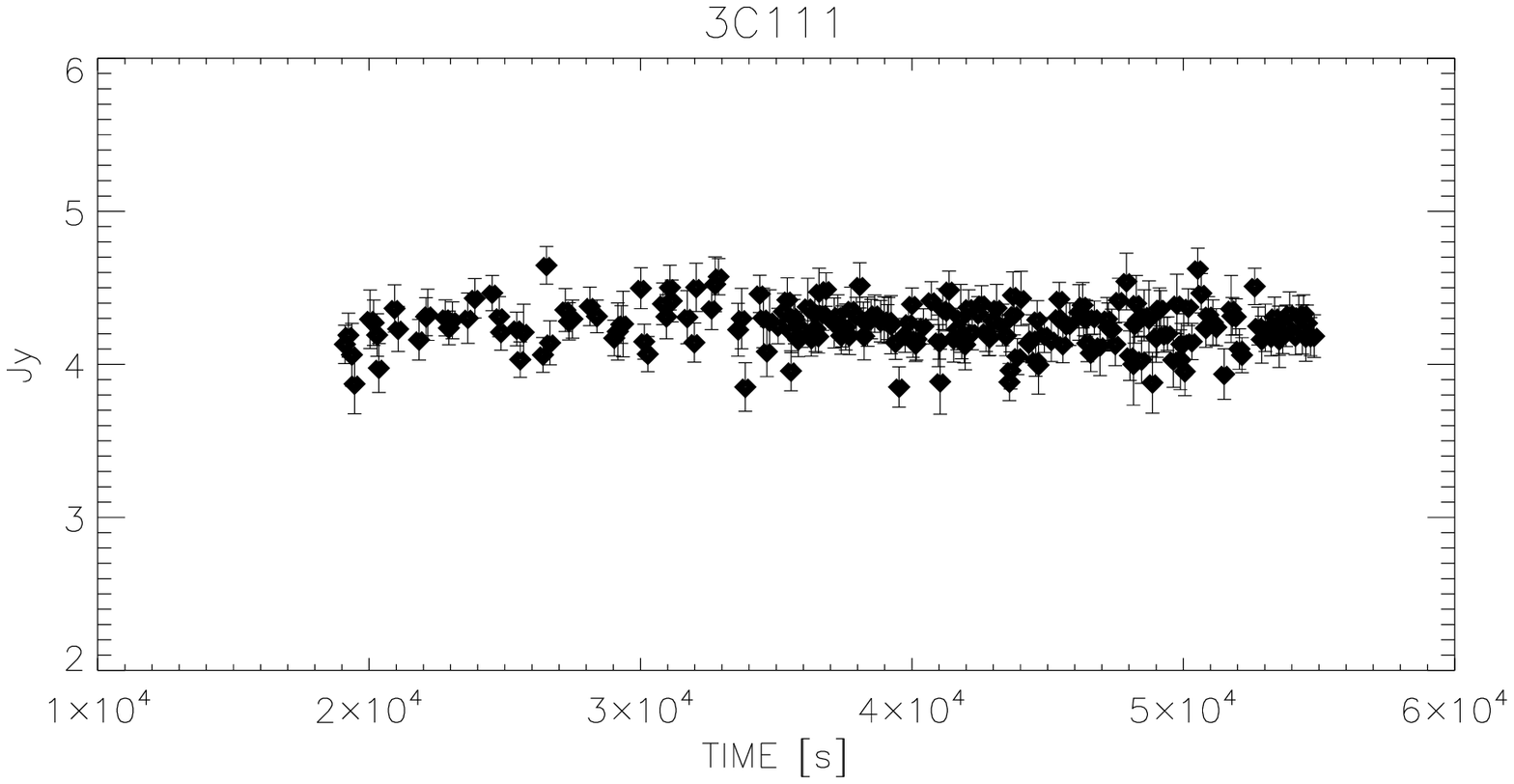} \hskip5mm 
\includegraphics[width=80mm]{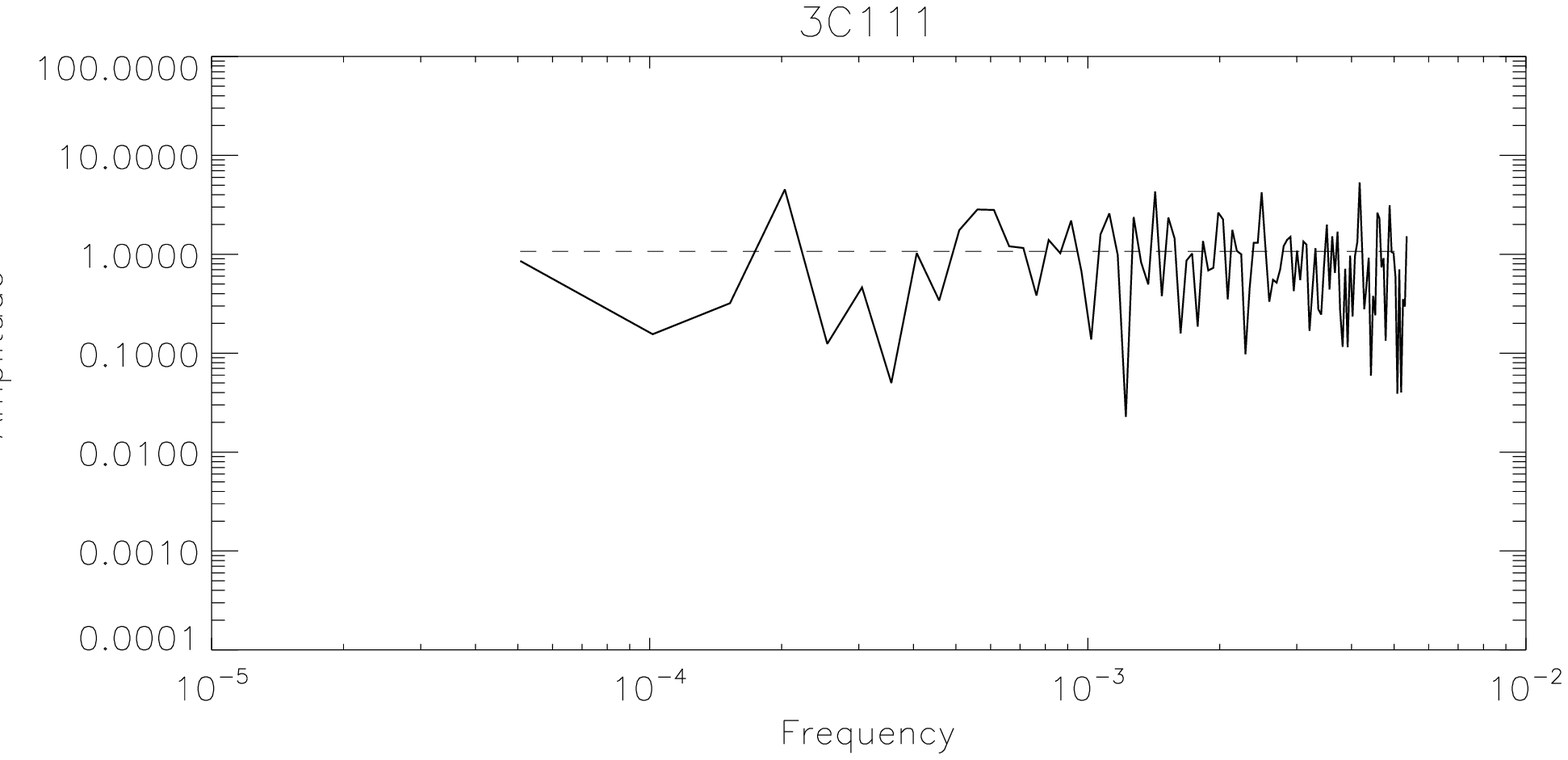}
\caption{The 22-GHz lightcurves (left panels) and Scargle periodogram (right panels) of 3C~111 observed at three consecutive days (from top to bottom) in February 2013. The sampling frequency of the periodograms is in units of s$^{-1}$. \label{fig:Feb3c111_22}}
\end{figure*}

\begin{figure*}[t!]
\centering
\includegraphics[width=80mm]{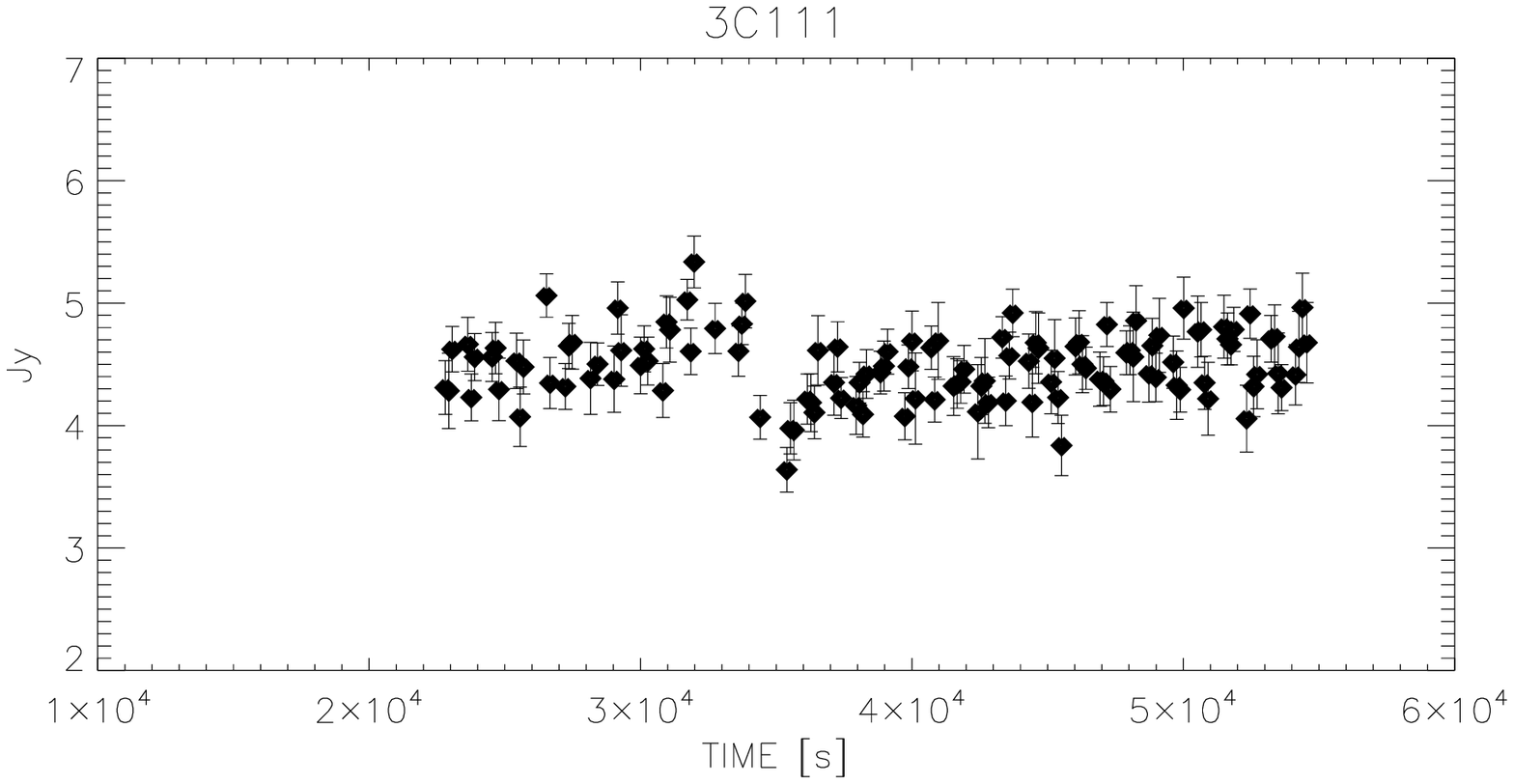} \hskip5mm 
\includegraphics[width=80mm]{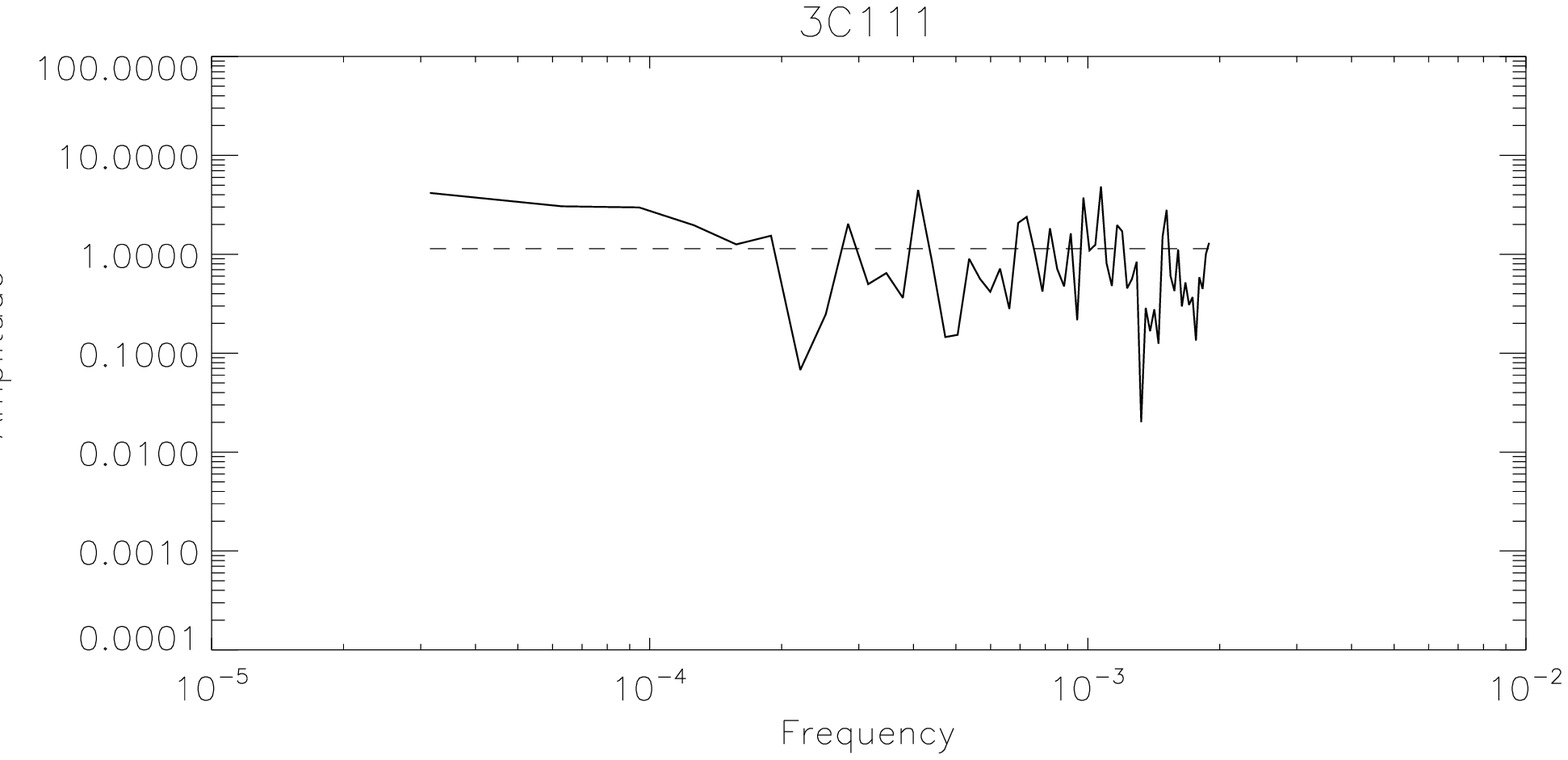} \\  
\includegraphics[width=80mm]{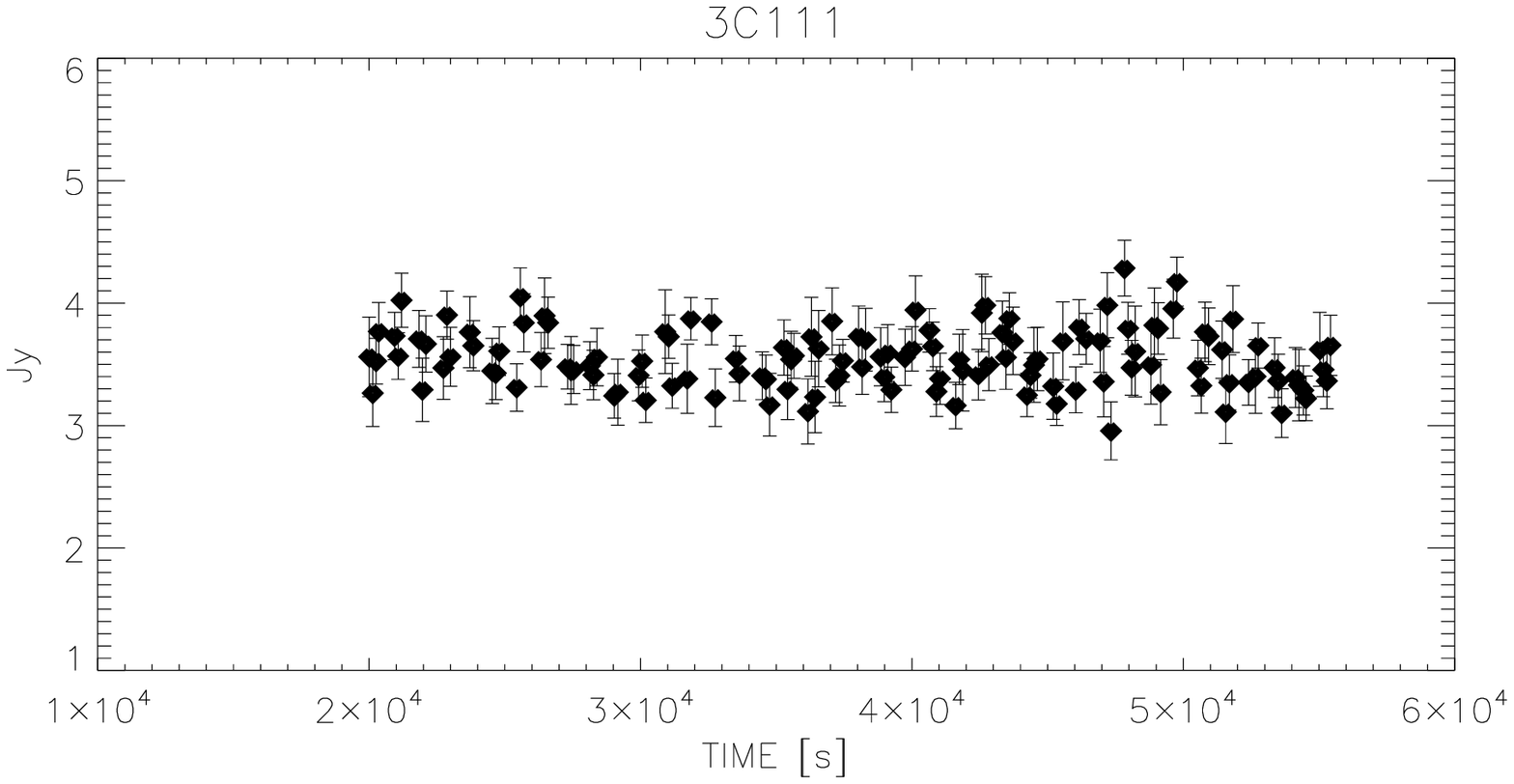} \hskip5mm 
\includegraphics[width=80mm]{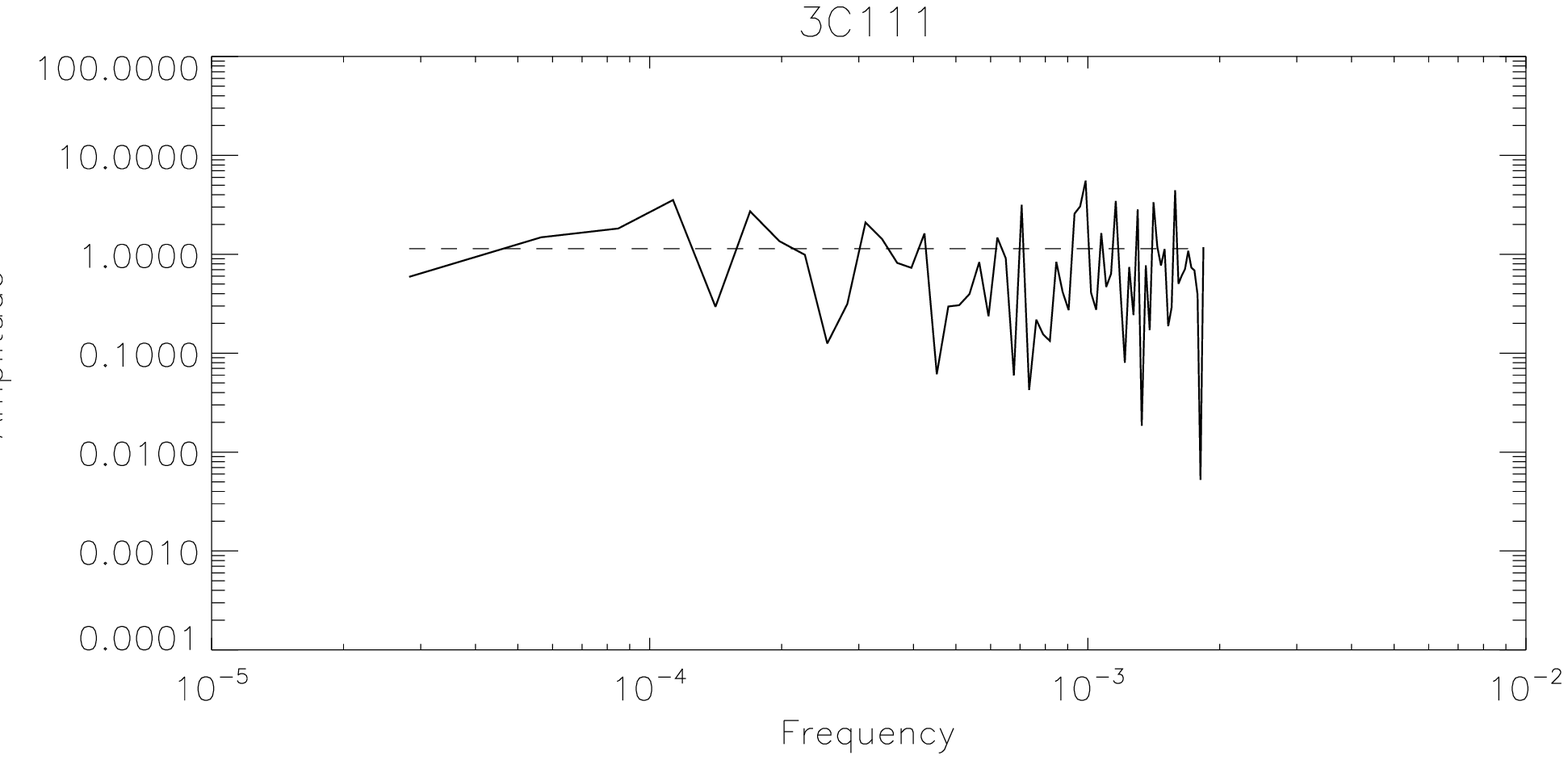} \\  
\includegraphics[width=80mm]{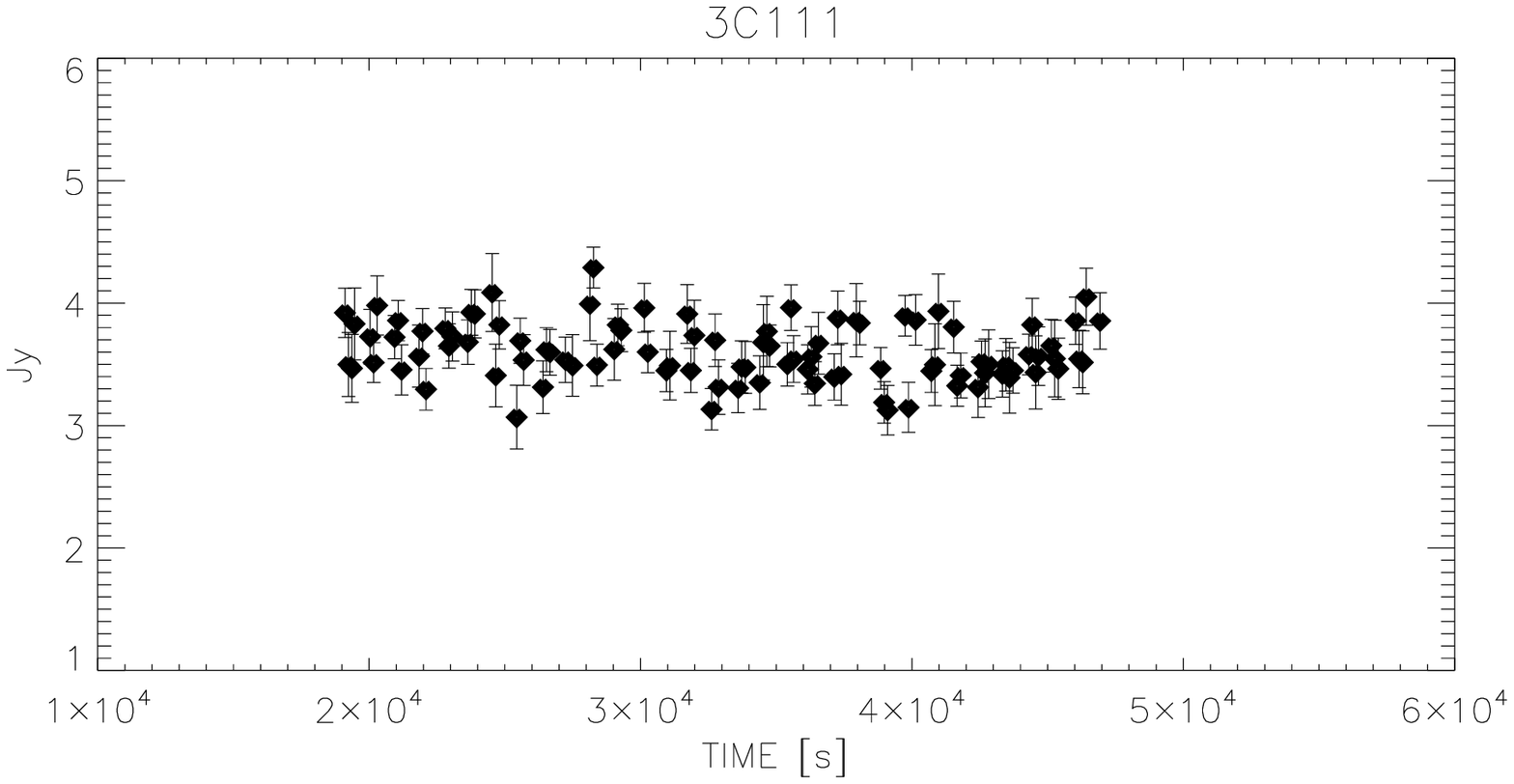} \hskip5mm 
\includegraphics[width=80mm]{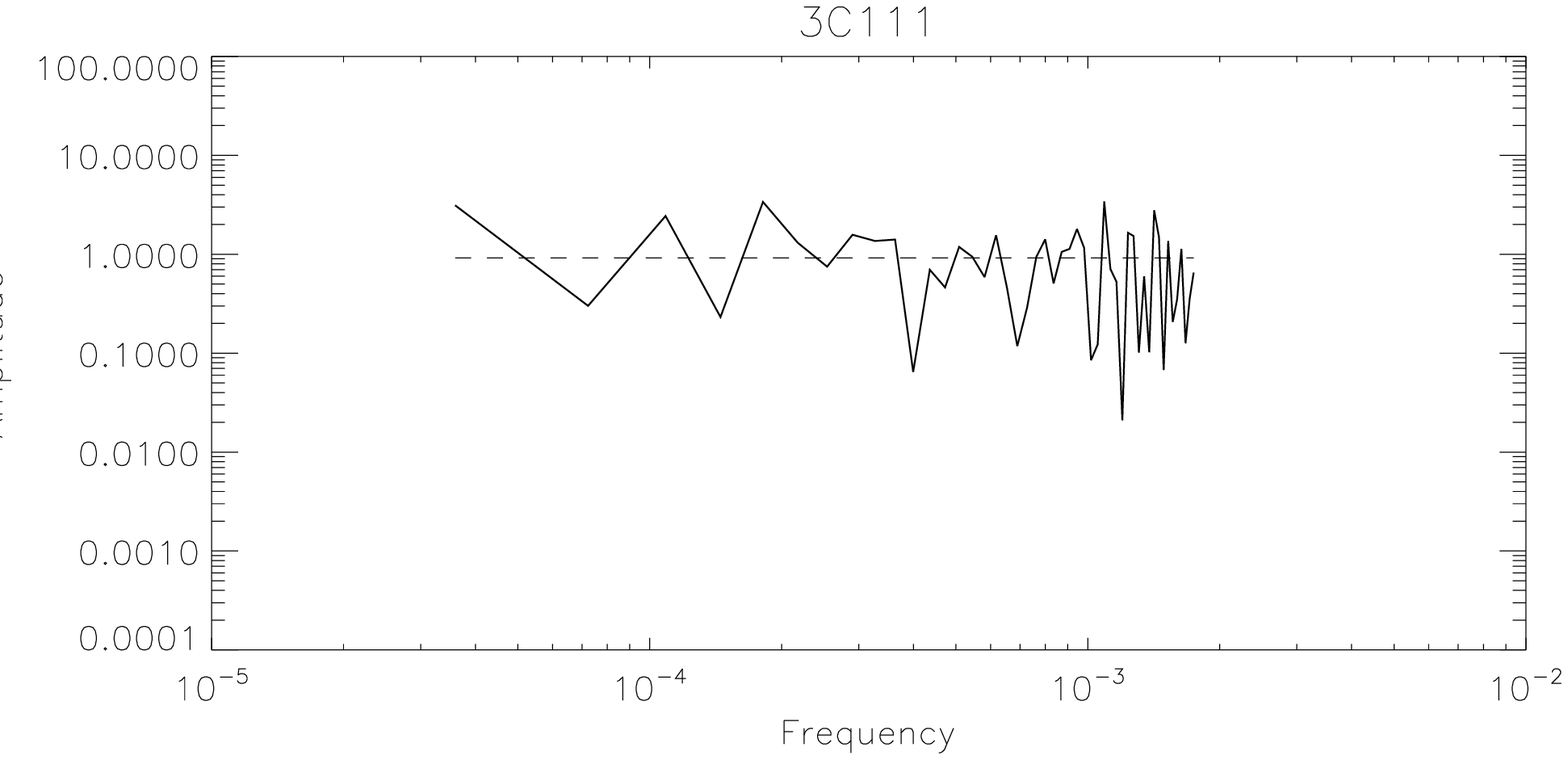}
\caption{The 43-GHz lightcurves (left panels) and Scargle periodogram (right panels) of 3C~111 observed at three consecutive days (from top to bottom) in February 2013. The sampling frequency of the periodograms is in units of s$^{-1}$. \label{fig:Feb3c111_43}}
\vspace{1.5em}
\end{figure*}

\begin{figure*}[t!]
\centering
\includegraphics[width=80mm]{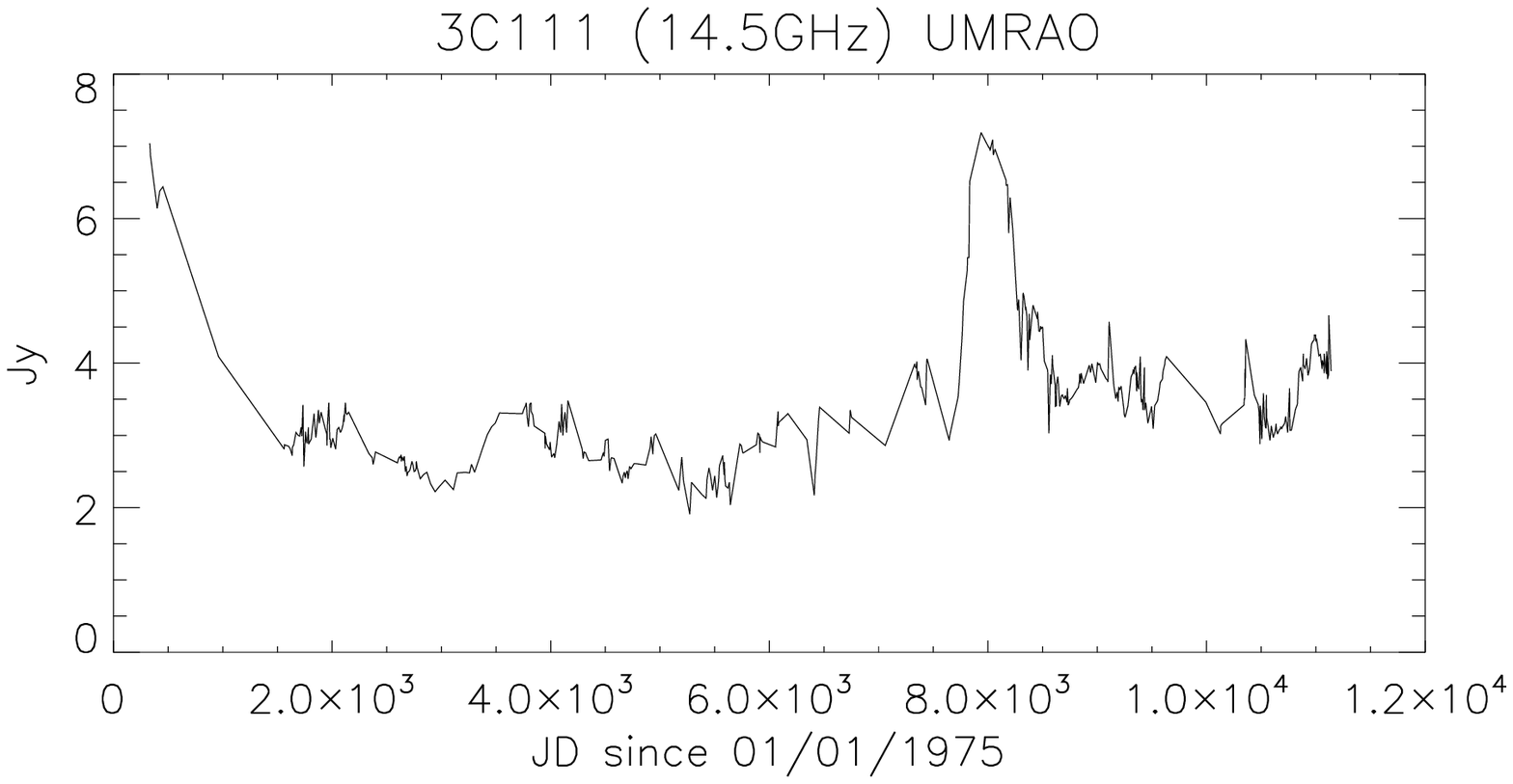} \hskip5mm 
\includegraphics[width=80mm]{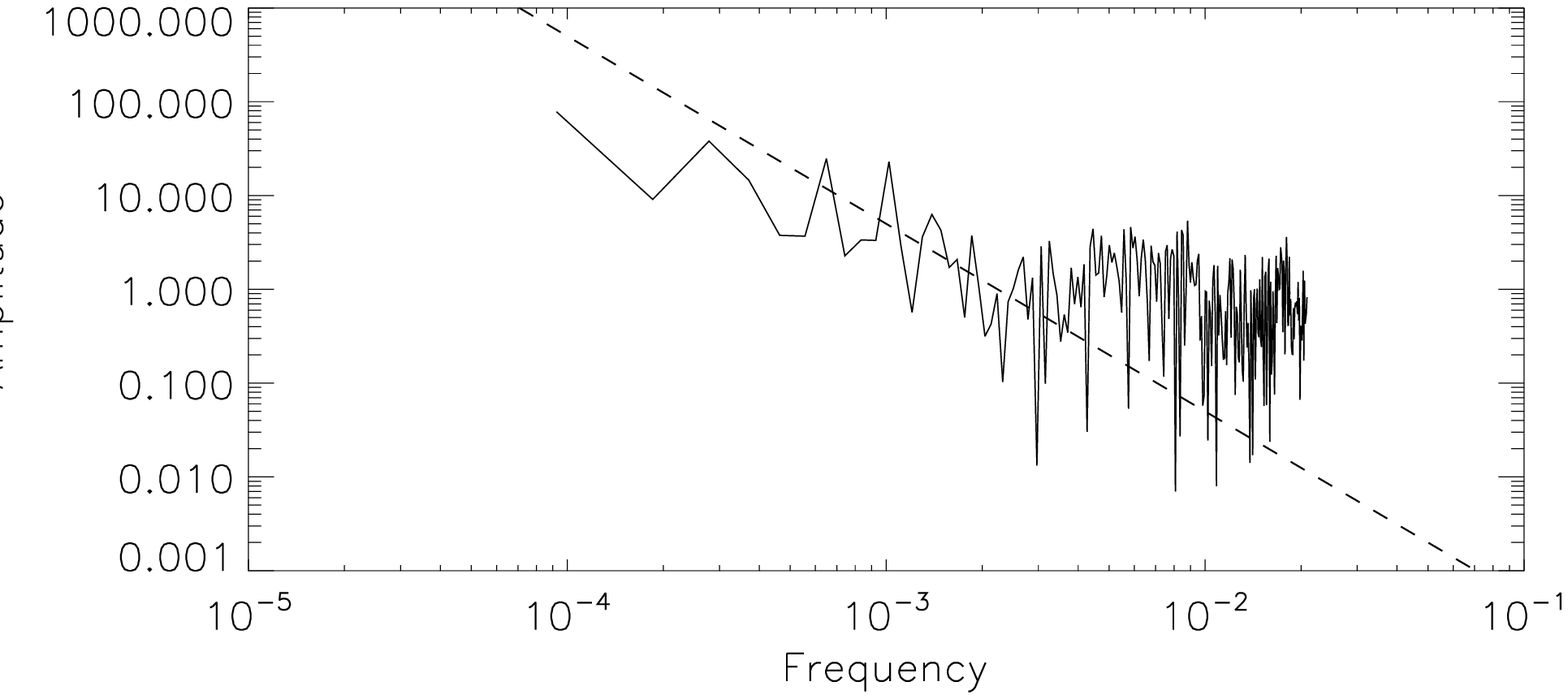} \\  
\includegraphics[width=80mm]{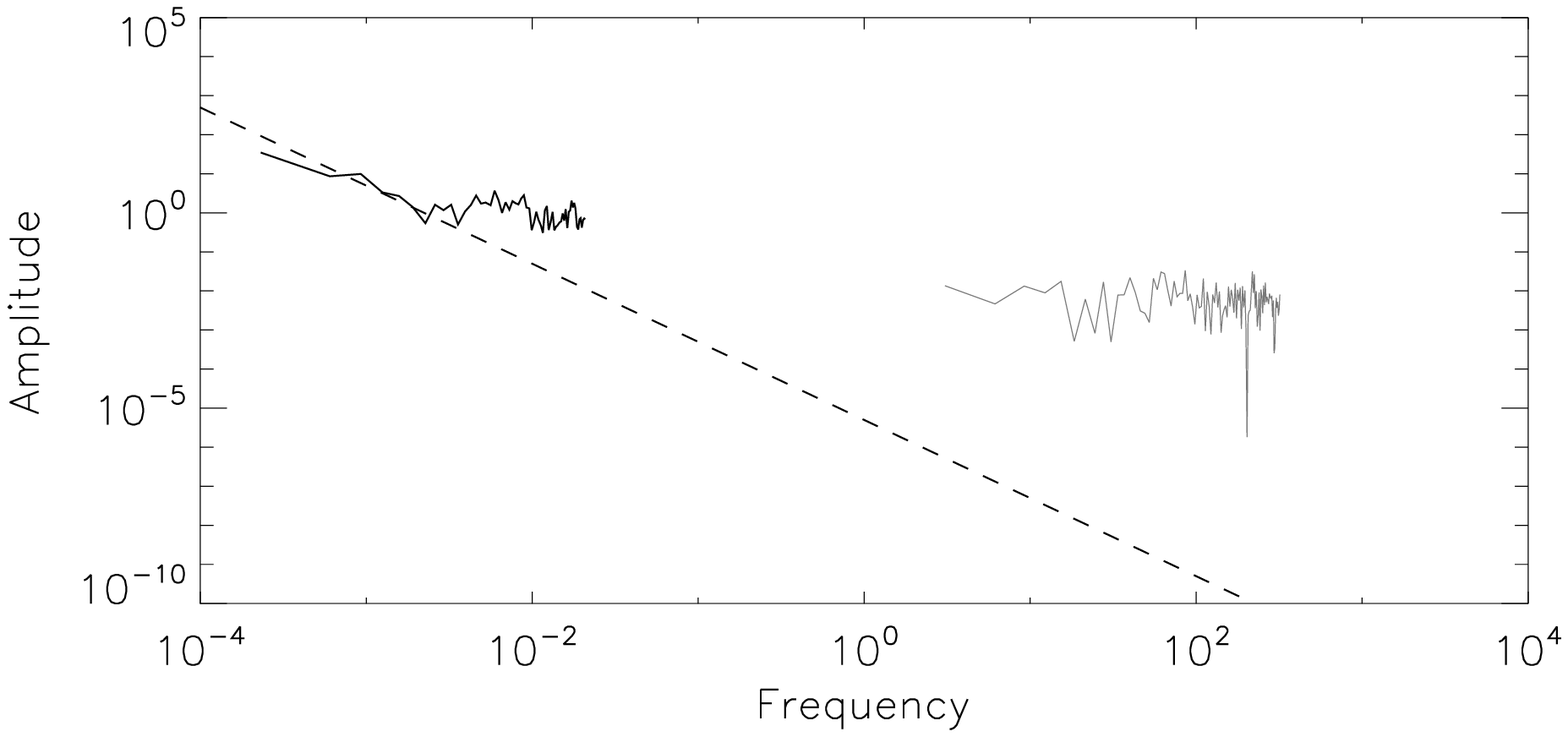} \hskip5mm 
\includegraphics[width=80mm]{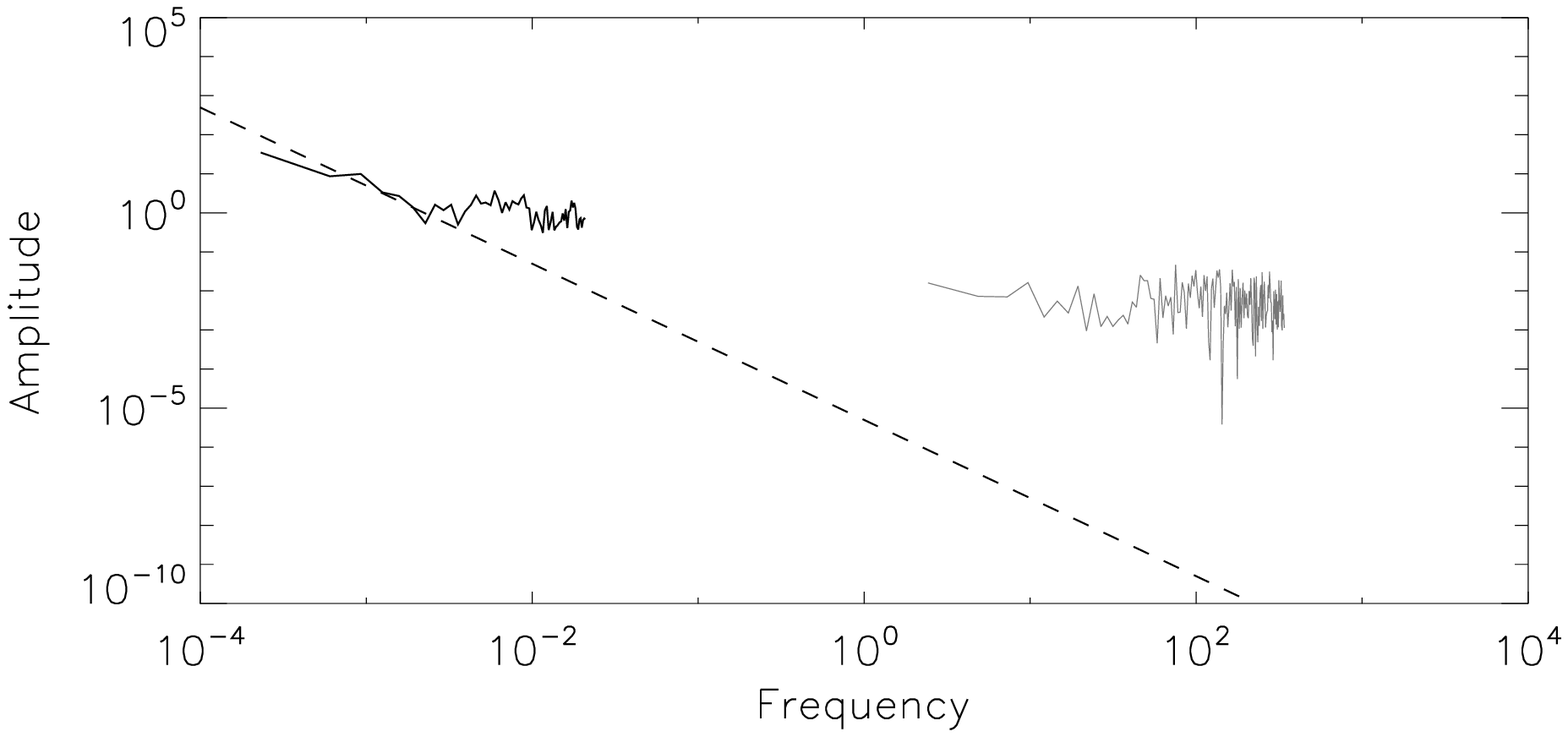}
\includegraphics[width=80mm]{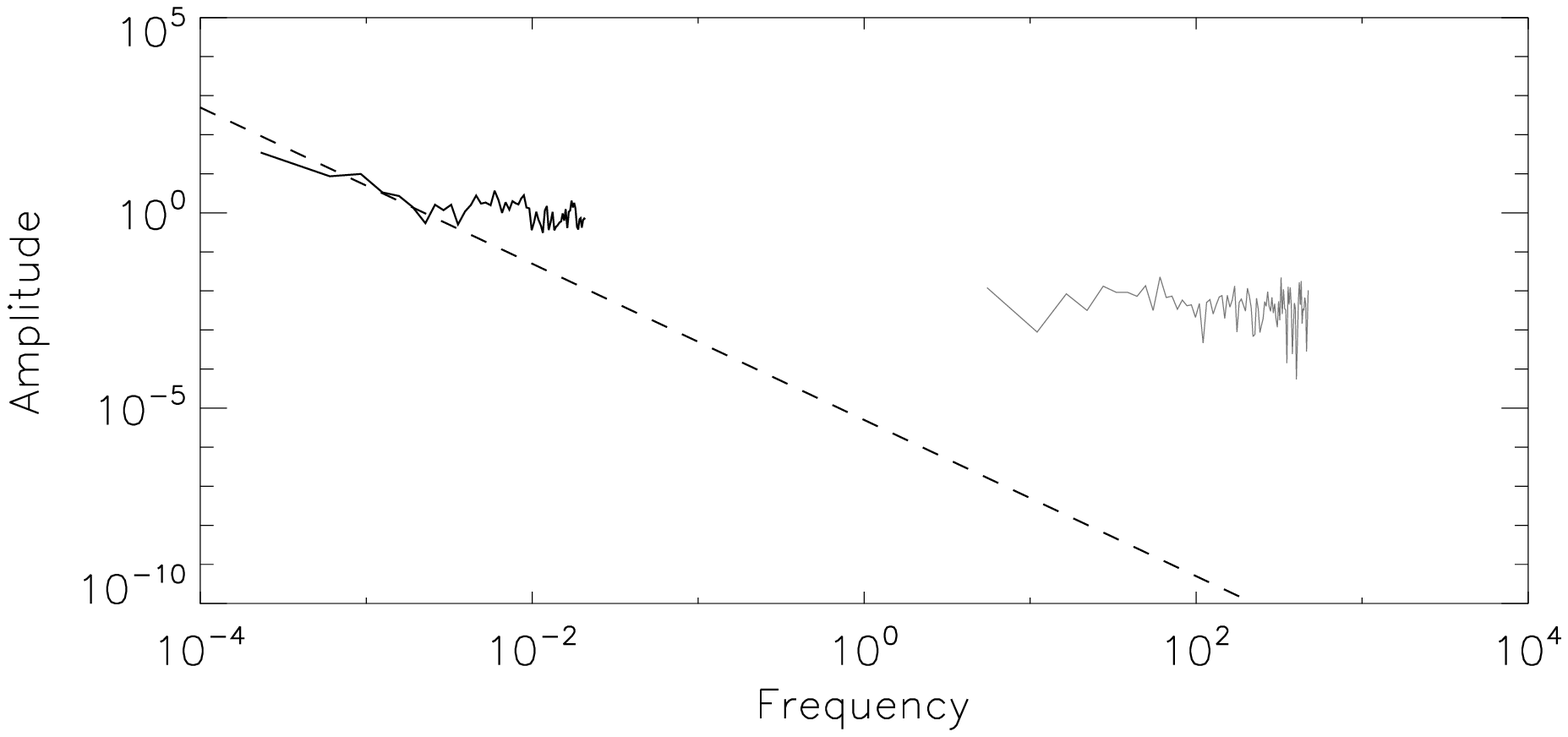}
\caption{\emph{Top left:} UMRAO lightcurve of 3C~111 observed at 14.5~GHz from 1975 to 2005. \emph{Top right:} The associated power spectrum; 
the dashed line indicates a theoretical random walk noise spectrum ($A_f \propto f^{\beta}$, $\beta=-2$). \emph{Center and bottom:} The Scargle periodogram from UMRAO data (black curve) along with periodograms from KVN 22-GHz data (gray curves) for three consecutive days in November 2012. The sampling frequency of the periodograms is in units of day$^{-1}$. Again, the dashed lines indicate theoretical random walk noise laws. \label{fig:3c111UMRAO}}
\end{figure*}

\begin{figure*}[t!]
\centering
\includegraphics[width=80mm]{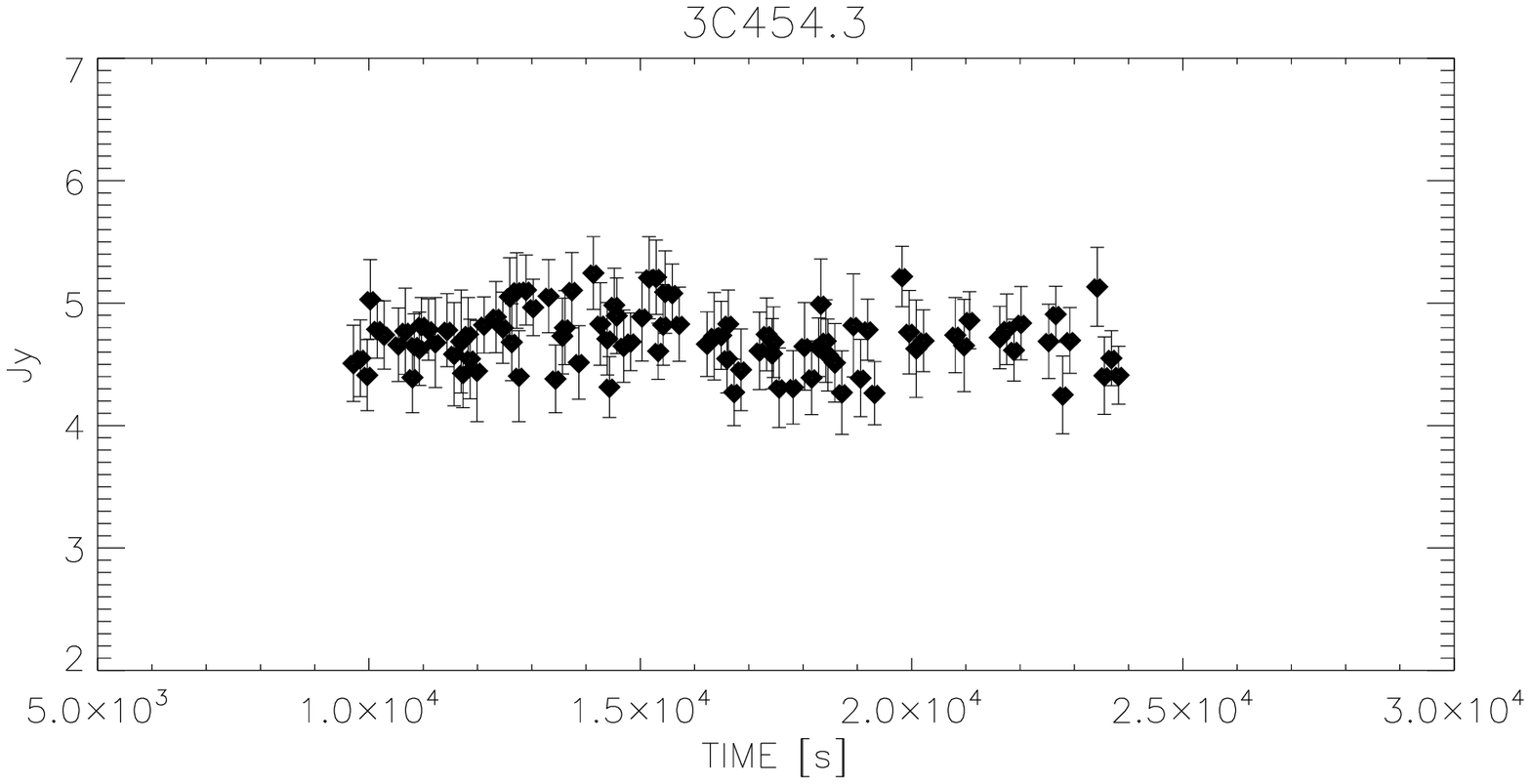} \hskip5mm 
\includegraphics[width=80mm]{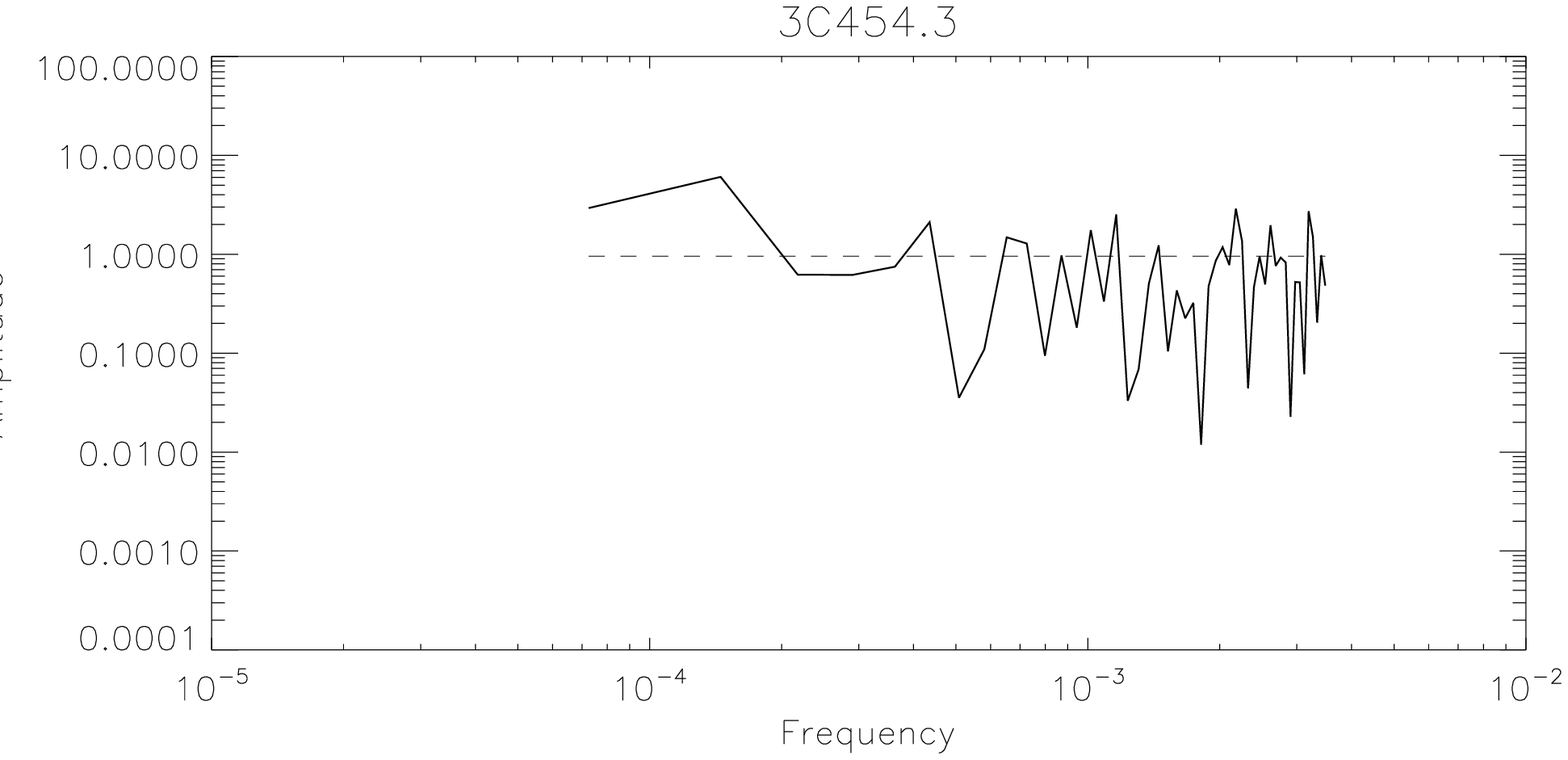} \\  
\includegraphics[width=80mm]{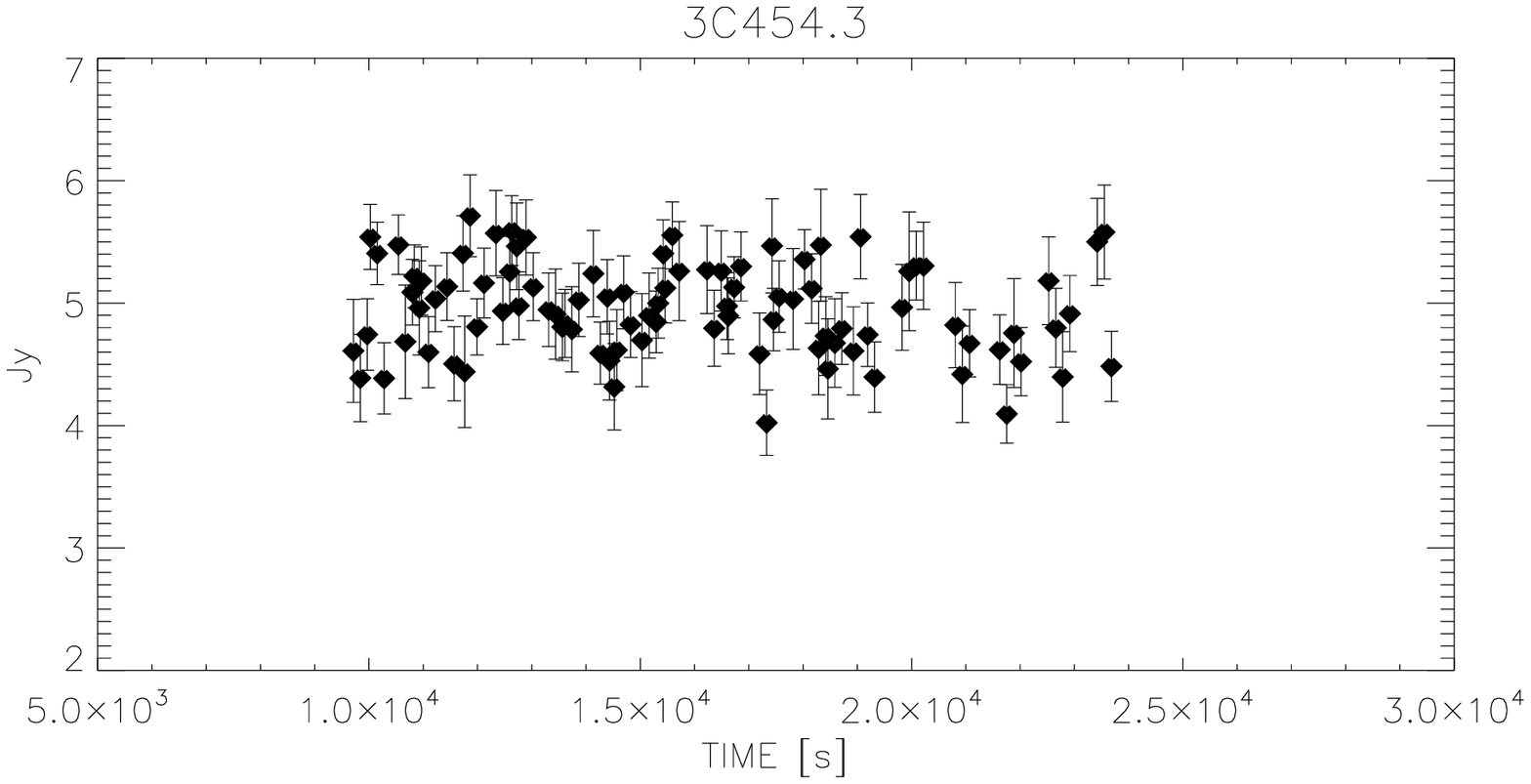} \hskip5mm 
\includegraphics[width=80mm]{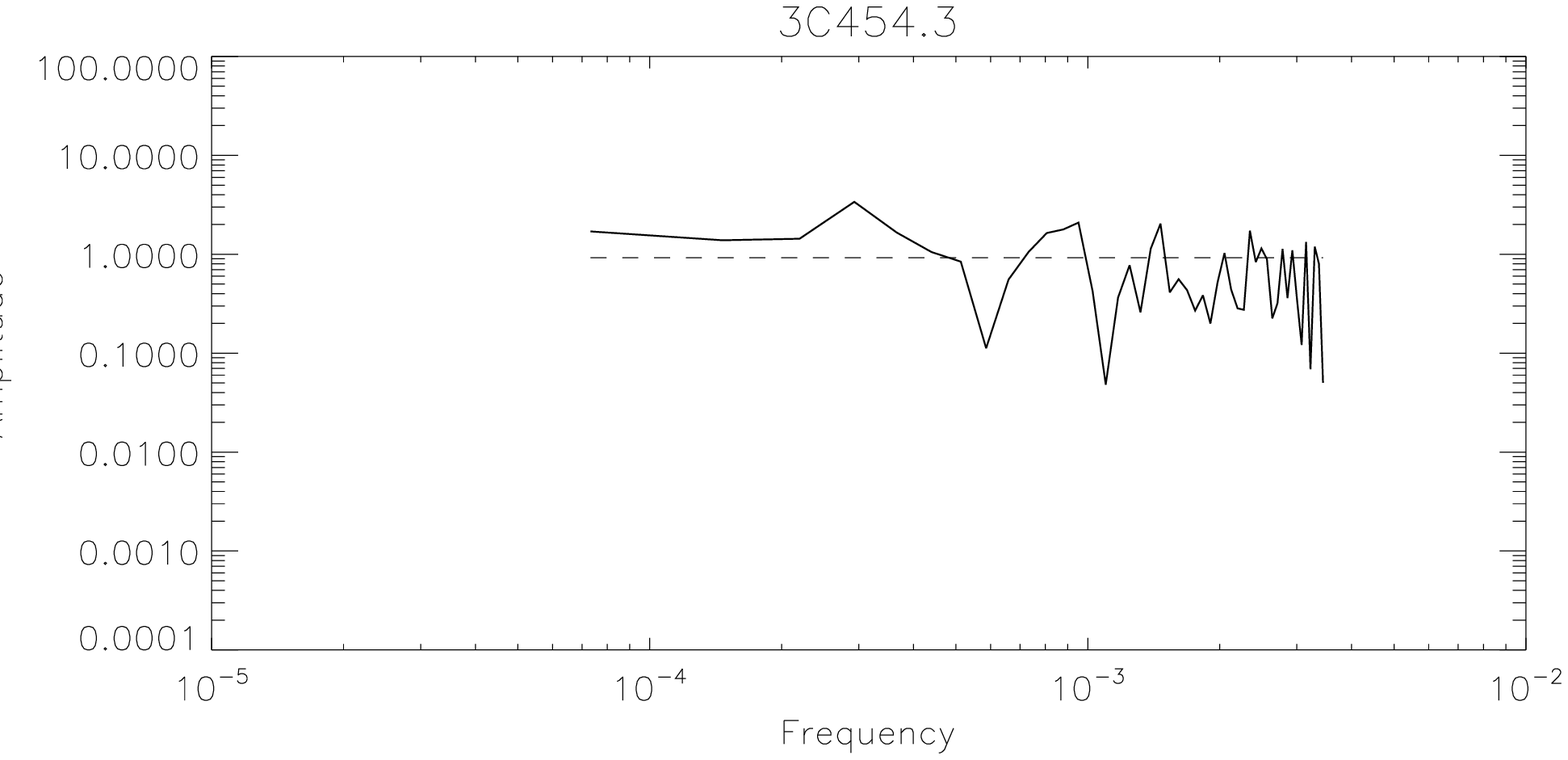}
\caption{Lightcurves (left panels) and Scargle periodograms (right panels) of 3C~454.3 observed in April 2013. \emph{Top:} 22~GHz. \emph{Bottom:} 43~GHz. The sampling frequency of the periodograms is in units of s$^{-1}$. \label{fig:3c454}}
\end{figure*}

\begin{figure*}[t!]
\centering
\includegraphics[width=80mm]{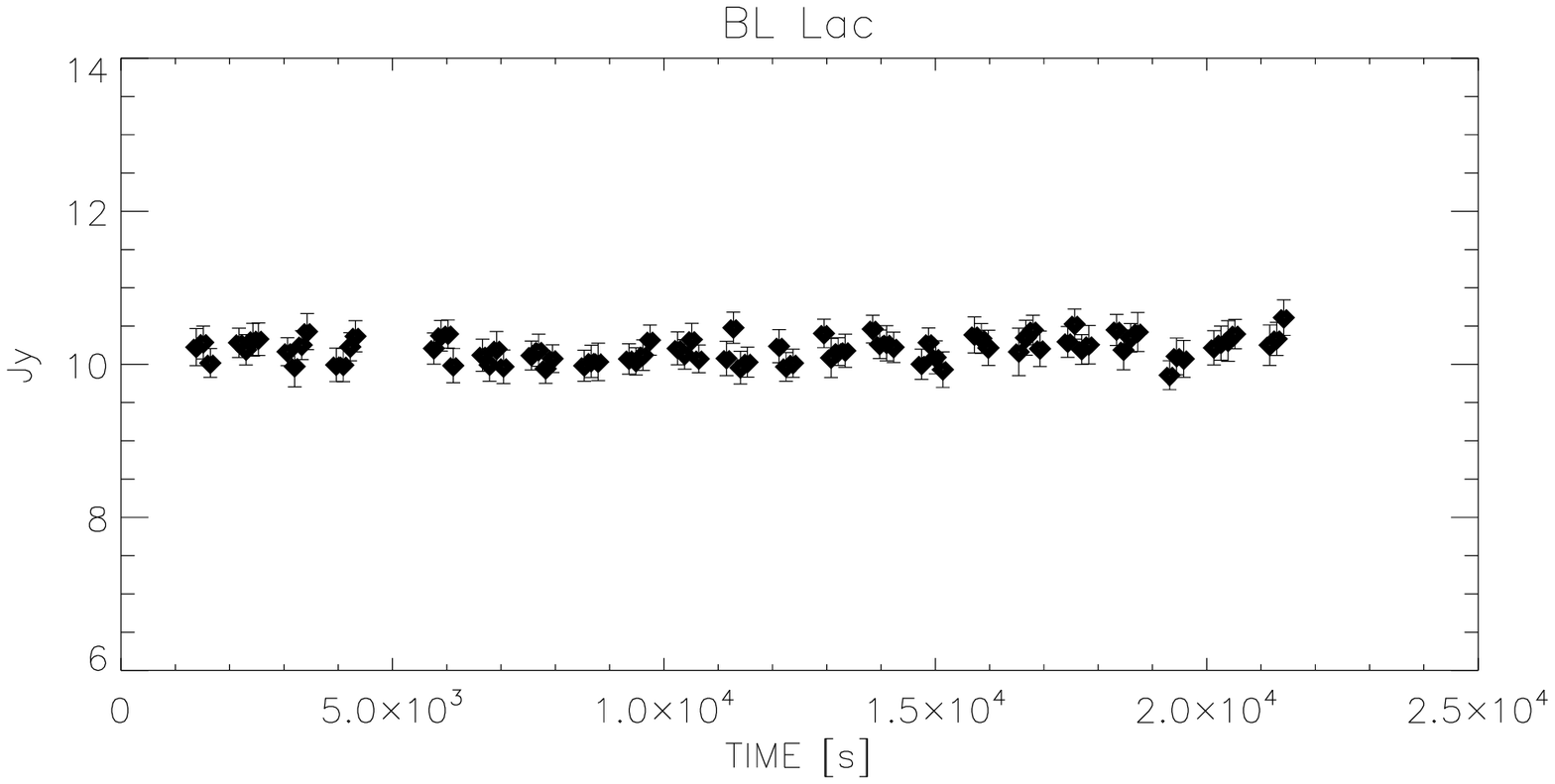} \hskip5mm 
\includegraphics[width=80mm]{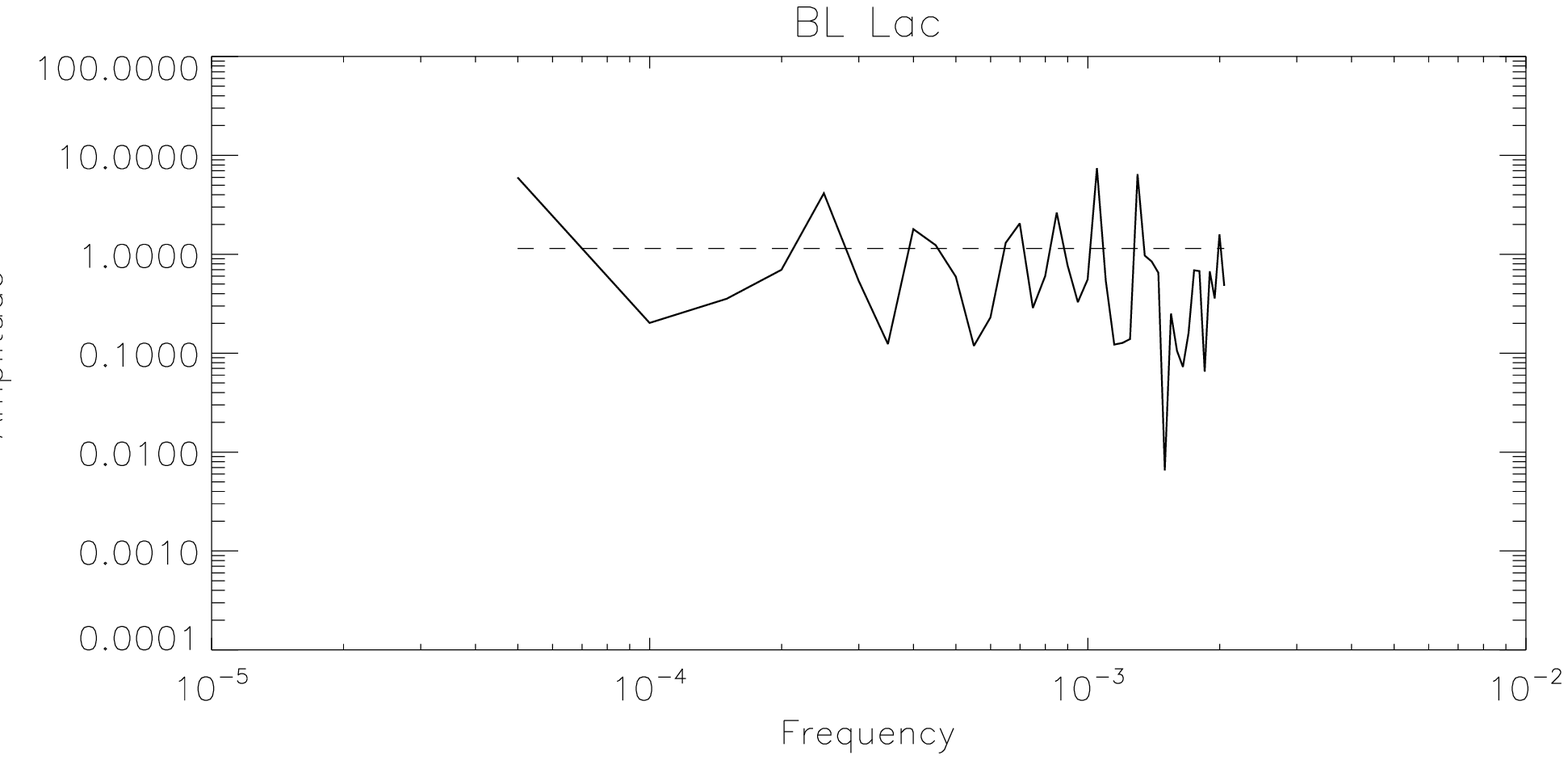} \\  
\includegraphics[width=80mm]{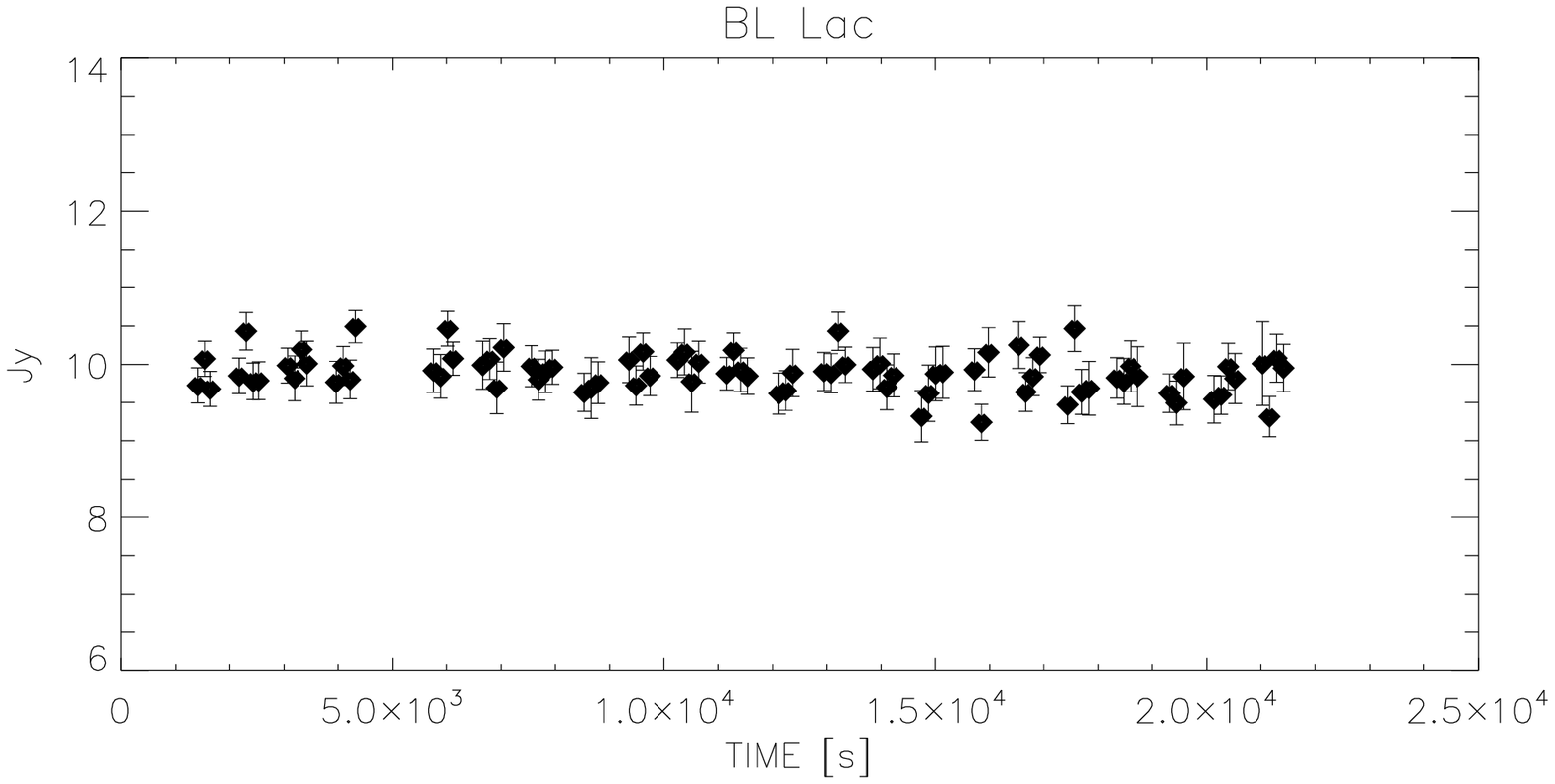} \hskip5mm 
\includegraphics[width=80mm]{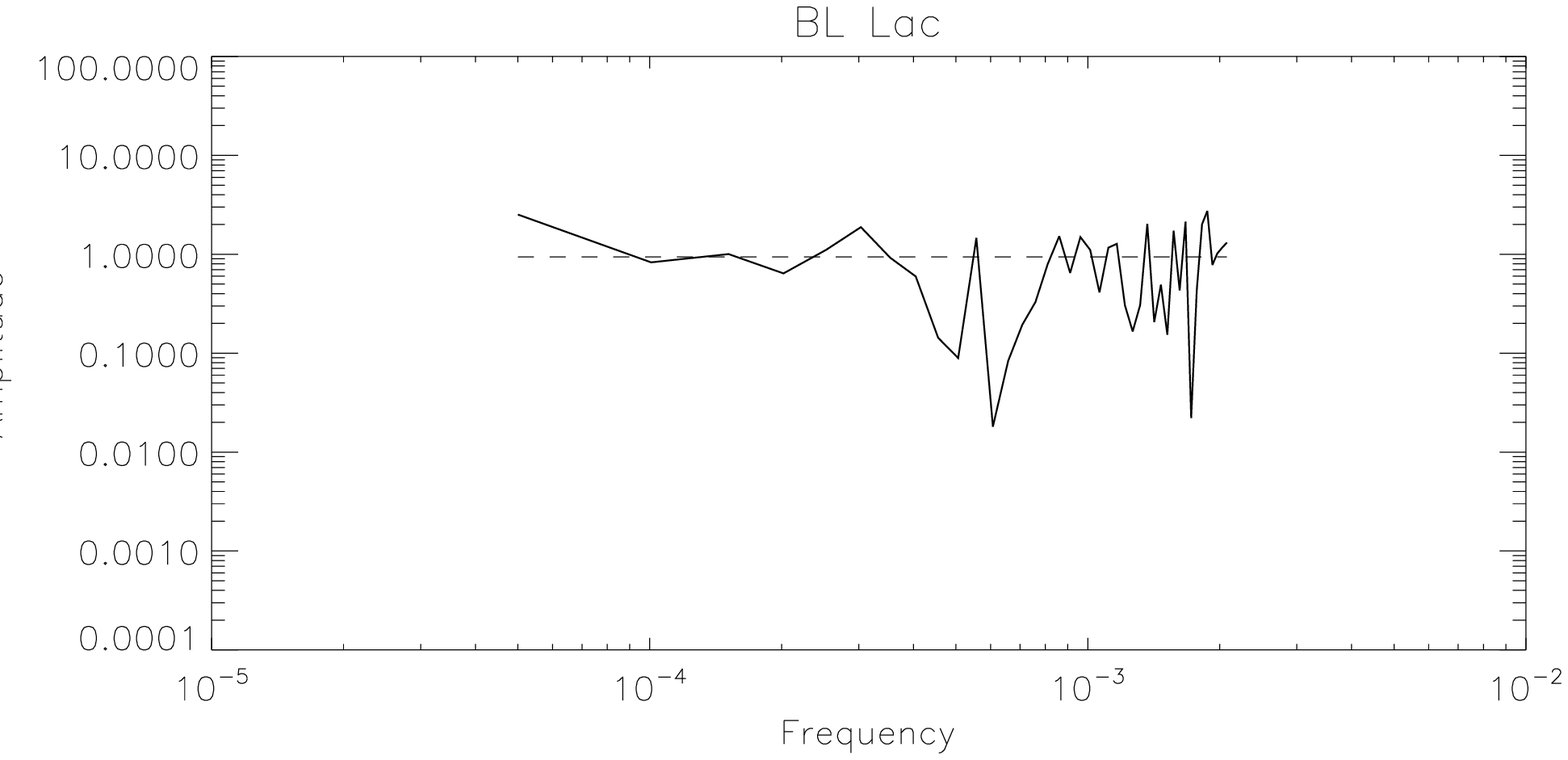}
\caption{Lightcurves (left panels) and power spectra (right panels) of BL Lacertae observed in April 2013. \emph{Top:} 22 GHz. \emph{Bottom:} 43 GHz. The sampling frequency of the periodograms is in units of s$^{-1}$. \label{fig:BLlac}}
\end{figure*}

Unfortunately, weather or technical conditions were difficult in general; we could achieve only four successful observation runs, out of 13 in total, resulting in high-quality data, mostly due to rain or cloudy sky. High frequency data were especially vulnerable to humidity; we obtained only one epoch of 86-GHz data and no data at 129~GHz. We had two epochs of three consecutive clear nights each in 2012 and 2013, resulting in long and high-quality light curves of 3C~111. At the low frequency bands of 22 and 43~GHz, 3C~454.3 and BL Lac could be observed in good conditions, however for one epoch only. In spring 2012, observations of 3C~279 with KVN Yonsei and Ulsan stations provided a long and densely sampled light curves at 86~GHz. In late 2013, we tried to catch other sources such as 4C~69.21 at higher frequencies of 86 and 129 GHz, only to fail mostly due to humid weather. Despite our efforts, we failed to combine the two light curves obtained from the paired stations for BL Lac (in April 2013), 3C~111 (in February 2013), and 3C~454.3 (in April 2013) because the quality of the two data sets was too different due to technical problems at one of the stations. For a detailed observations journal, see Tables~\ref{tab:obsstatus1} and \ref{tab:obsstatus2}. In some cases, data that appeared to be of high quality (in terms of good agreement between cross-scan source profile and theoretical Gaussian) turned out to suffer from systematic uncertainties caused by technical glitches like incorrect time resolution, receiver problems, or further technical errors at a KVN station; those data were excluded from the analysis. Data quality classification is discussed in the following section.

\section{Data Analysis\label{sec:data}}

The primary data extraction and calibration made use of the Gildas CLASS software package.\footnote{\url{http://www.iram.fr/IRAMFR/GILDAS}} Each scan across a target provided a (Gaussian for point sources, as was the case for our observations) profile of the amplitude as function of position angle. We sorted out bad scans according to the clearness of the Gaussian shape of the amplitude profile; we assigned a quality flag ``high'' to scans showing a clear Gaussian profile, and ``low'' to scans showing nothing like a Gaussian (see Tables~\ref{tab:obsstatus1} and \ref{tab:obsstatus2}).

      We fit a linear baseline to each amplitude profile in order to remove (instrumental and sky) background emission. We fit a Gaussian curve to the amplitude and used the best-fit peak value of the profile as one flux data point. By combining four consecutive data points from one full scanning sequence, we obtained one flux measurement point for a light curve, using the mean value as flux measurement value.
We used the standard deviation of the four scan values as the initial estimate of the statistical error of a given flux measurement. However, error bars obtained this way are seriously affected by low number statistics. Robust error bars can be found from an iterative procedure:\footnote{Iterative error estimates are well established in stellar dynamics, especially for calculating velocity dispersions from stellar velocities with poorly constrained uncertainties (see \citealt{armandroff1986,hargreaves1994} for detailed discussion).} (1) calculate a first error estimate $\sigma_{0,i}$ for data point $i=1, 2,...$ from the standard deviation of four scan values; (2) compute the average $\sigma_{0,i}$, i.e. $\langle\sigma_{0,i}\rangle$, for all data points of a given light curve; and (3) compute the final error $\sigma_i$ for a given data point via
\begin{equation}
\label{eq:sigma}
\sigma_i^2 =\frac{1}{2} \sigma_{0,i}^2 + \frac{1}{2} \langle\sigma_{0,i}\rangle^2 
\end{equation}

	We removed data points deviating more than 3$\sigma$ from the mean value. In general, data obtained at the beginning and finishing parts of the observation have large error bars caused by the low elevation of the target and were removed. Furthermore, the KVN antennas seemed to have difficulties in pointing at a target around the time of its transit, resulting in the loss of about 10 to 15 minutes of data in the middle of the light curves obtained in November 2012.
	
	The light curves of the calibrators were normalized to unity averages and were fit with 2nd order polynomials using IDL.\footnote{Interactive Data Language, ITT Exelis Inc., McLean (Virginia)} The uncalibrated light curves of the targets were then divided by the calibrator model curves, resulting in amplitude calibrated lightcurves of the targets. The maximum elevation difference between target and calibrator was 20$^{\circ}$. According to the KVN gain curve,\footnote{\url{kvn.kasi.re.kr/status_report_2013/gain_curve.html}} this difference corresponds to about 3 $\%$ difference in normalized gain at 22 and 43 GHz, and up to about 6\% at 86 GHz.

	We achieved a typical time resolution of three minutes and, occasionally, one minute whenever the light curves from paired KVN stations could be combined. For some epochs, only parts of the light curves could be combined since the data from one of the stations were usable only for a part of the total observing time. In some cases (and even when the observation conditions were good at both KVN sites), we found a systematic offset between the two light curves. The most obvious case occurred in the data for 3C~454.3 obtained in April 2013 with a relative offset value of 6.8$\%$. Assuming this was an effect occurring at one of the stations (specifically weather, gain calibration errors, or incorrect aperture efficiency estimates), we multiplied the ratio of the averages to one light curve to remove the offset before combination.

	Flux densities were converted from antenna temperature (in Kelvin) to flux density (in Jansky) using
\begin{equation}
{\rm [Jy/K]} = \frac{2k}{A_{eff}}=\frac{2k}{\eta \pi r^2},
\end{equation}
where $k$ is the Boltzmann constant, $A_{eff}$ is the effective antenna aperture, $\eta$ is the aperture efficiency, and $r$ is the antenna radius.
The conversion factors are different depending on the sites, frequencies, and time. For KVN, we found values ranging from 12.4 to 14.2 throughout our observation period.

	To analyze the variability of the light curves, we applied modulation index ($m$, in units of \%) analysis, using \citep{Kraus2003}
\begin{equation}
\label{mindex}	
	m = \frac{stddev}{mean} \times 100 \% 
\end{equation}	
where $stddev$ and $mean$ are the standard deviation and mean values of the flux densities. 

	We also performed a reduced $\chi^2$ test assuming a constant flux density with \citep{Kraus2003}
\begin{equation}
\label{chisq}
	\chi^2_r = {{1} \over {N-1}} \sum_{i=1}^{N} {{(S_i - \langle S\rangle)^2} \over  {\sigma_i^2} }
\end{equation}
where $N$ is the number of data points, $S$ is the flux density, $\sigma_i$ is the error derived according to Equation~(\ref{eq:sigma}), and $\langle\cdot\rangle$ denotes the average of the enclosed quantity.
	The modulation indices and reduced $\chi^2$ values are listed in Table~\ref{tab:param}.

Taking the standard deviation of a given lightcurve and the observed time period as proxies for the amplitude of flux variation and the variability time scale, respectively, we obtained limits on the brightness temperatures using \citep{Wagner1995},
\begin{equation}
\label{Tb}
T_{b}=4.5 \times {10^{10}} A_{var} \bigg[\frac{\lambda d}{t_{var}(1+z)}\bigg]^2
\end{equation}
where $A_{var}$ is the amplitude of flux variation (in units of Jy), $\lambda$ is the observation wavelength (in cm), $d$ is the distance to the source (in Mpc), $t_{var}$ is the variability time scale (in days), and $z$ is the redshift of the source.
Our upper limits on brightness temperatures are listed in Table~\ref{tab:param}. They range from $1.2 \times 10^{15}$\,K to $2.9 \times 10^{18}$\,K, much higher than the inverse Compton limit of $T_{max}=10^{12}$\,K. In general \citep[e.g.,][]{Fuhrmann2008}, the occurrence of brightness temperatures in excess of the inverse Compton limit is interpreted as an effect of Doppler boosting. The corresponding Doppler factors can be found from \citep{Wajima2014},
\begin{equation}
\label{d_factor}
\delta = (1+z)\bigg(\frac{T_{b}}{10^{12} K}\bigg)^{1/(3-\alpha)}
\end{equation}
where $z$ is the redshift of the source and $\alpha$ is the spectral index which is typically assumed to be $0.6$, defined via $S_{\nu}\propto\nu^{-\alpha}$ \citep[cf.,][]{KimJY2013}.

The emission from AGN is known to follow red-noise statistics with amplitudes $A_f\propto f^{\beta}$ ($f$ being the sampling frequency) with $\beta<0$ \citep[e.g.,][]{Park2014}. Accordingly, we used Scargle periodogram analysis \citep{Scargle1982} to constrain the statistics of our lightcurves. Especially, we aimed at distinguishing intrinsic variability (which is likely to show $\beta<0$) from instrumental and atmospheric fluctuations which would show white noise behavior ($\beta=0$).

\section{Results}

\subsection{3C 279}
Being our brightest target source, 3C279 was observed in May 2012, providing the only 86-GHz data set we could obtain (Figure~\ref{fig:3c279}). The two paired KVN stations contributed a dense and high-quality lightcurve throughout the whole observation period. The lightcurve appears flat within errors with slight ups and downs at the beginning and the end; these deviations come with relatively large error bars and scatter, and occur at times of low elevations. The power spectrum is essentially flat (powerlaw index $\beta\approx0$), implying a white noise time series.

\subsection{3C 111}
Partially thanks to its high declination, 3C~111 could be observed for the longest observation time among our targets, 56 hours in total. 
In November 2012 (Figure~\ref{fig:3c111Nov}), the light curves of all three observing days show remarkably constant flux density of around 2.8~Jy, with $m$ values lingering around 4$\%$, and reduced $\chi^2$ values around 1.3. The power spectra are flat.

3C~111 still seemed to be equally calm three months later, when observed at 22~GHz in February 2013 (Figure~\ref{fig:Feb3c111_22}). The weather conditions were consistently fine, the light curves are almost flat with mean fluxes around 4.3~Jy. The $m$ values are 2.8\%, 3.0\%, and 3.4\%, respectively. Reduced $\chi^2$ values range from 1.1 to 1.3. At 43~GHz (Figure~\ref{fig:Feb3c111_43}), the light curves show larger error bars than the 22~GHz data and show slight systematic variations. Both reduced $\chi^2$ and $m$ values are a little higher than those of the 22~GHz lightcurves. The data from the first day show a "break" feature caused by an antenna pointing instability around the time of the transit of the target. The power spectra for both frequencies are almost constant. 

For a deeper statistical analysis, we used 14.5-GHz long-term flux monitoring data taken from 1975 to 2005 by the University of Michigan Radio Astronomy Observatory (UMRAO) for comparison with our 22-GHz data sets \citep[the UMRAO data are discussed in detail in][]{Park2014}. We calculated Scargle periodograms for all time series and normalized the power spectra from UMRAO and KVN data consistently in order to make them compatible; the results are shown in Figure~\ref{fig:3c111UMRAO}. We compare the observed power spectra to a theoretical random walk noise ($\beta=-2$) law which appears to be a generic feature of blazar power spectra \citep{Park2014}; the flat tail of the UMRAO power spectrum is due to known sampling effects. As it turns out, all KVN power spectra are located above the theoretical line by several orders of magnitude.

\subsection{3C 454.3}
In April 2013, 3C~454.3 provided moderately sampled light curves (Figure~\ref{fig:3c454}) with relatively large error bars at both 22 and 43 GHz. The $m$ and reduced $\chi^2$ values are comparable to the 43~GHz data for  3C~111 of February 2013. With short observing times and large scatter, the lightcurves lead to high upper limits on brightness temperature.

\subsection{BL Lac}
BL Lacertae was observed on the same day as 3C~454.3 for a somewhat longer time. Though the sampling of the light curves at 22 and 43~GHz is only moderately dense, the light curves are the flattest among all our sources: reduced $\chi^2$ values are 0.6 and 0.9, $m$ indices are 1.6\% and 2.5\%, respectively.

\section{Discussion}

Despite our efforts and a high time resolution of few minutes, we have not found indication for distinct intra-day variability in our target sources. On time scales from minutes to hours, the source fluxes did not change more than a few percent, the light curves are flat. 

The modulation indices range from 1.6\% to 7.6\% depending on source and epoch. These values are comparable to or slightly higher than those in the work by \citet{Kraus2003} or \citet{Gupta2012} in which the authors searched for and found, in part, variability in compact extragalactic radio sources. In our case, our relatively large $m$ values are caused mostly by the photometric uncertainty. The lightcurves are characterized by scatter rather than systematic deviations from the mean, as expected for the case of source intrinsic variability.

Our $\chi^2_r$ tests find values ranging from 0.6 to 1.8. These values are in agreement with no  variation (for significance levels given by $\chi^2_r\approx1.4$ for a false alarm probability of 0.1\%) in our lightcurves\footnote{Calculated using the Internet $\chi^2$ calculator provided by\\  \url{www.fourmilab.ch/rpkp/experiments/analysis/chiCalc.html}} (see also Table~\ref{tab:param}) -- with the exception of one 43-GHz lightcurve of 3C~111 obtained in February 2013 which shows a fluctuation that is actually caused by an antenna pointing instability during the transit of the target.
Overall, our lightcurves are in agreement with being flat (within errors). 

Even so, we were able to identify upper limits for flux variability and, at least formal, upper limits on brightness temperatures.
As we did not find intrinsic variability, we used the scatter in a given lightcurve and its length in time as proxies for the amplitude of flux variations and the variability time scale. The resulting, rather loose, upper limits on brightness temperatures ($10^{13}$ to $10^{18}$ K) exceed the Inverse Compton limit ($10^{12}$ K) \citep{Kellermann_Pauliny-Toth1969} as well as the equipartition limit ($3 \times 10^{11}$ K) \citep{Readhead1994} by two to six orders of magnitude.
Physically, we are dealing with brightness temperatures in the \emph{observer} frame, not with the intrinsic plasma temperatures in the \emph{emitter} frame. Relativistic Doppler boosting leads to an upscaling of the \emph{apparent} brightness temperature by the appropriate Doppler factor as given in Equation~(\ref{d_factor}). Accordingly, our upper limits on apparent brightness temperatures can be translated into (again, rather loose) upper limits on Doppler factors ranging from about 20 to about 900.

This view is supported by the results of the periodogram analysis. Overall, all power spectra are consistent with white noise statistics ($\beta=0$), in contrast to the known red noise behavior of AGN emission.
From the power spectrum of 3C~111 we see that the KVN power spectra of three different days lie several orders of magnitude above the theoretical power-law line ($A_f \propto f^{-2}$). This suggests (again) that our data are dominated by measurement noise and provide upper limits on the intrinsic variability only, which would have to be at sub-per cent level. We note however that it is not clear if the $f^{-2}$ law is valid on time scales of minutes. 

In our analysis, we have not followed flux variations on time scales longer than hours. 3C~279 had a high flux density of 28 Jy. 3C~111 showed different flux densities in November 2012 (2.8 Jy) and February 2013 (4.2 Jy), but we were not able to follow the trend since our observations covered two epochs only. This suggests that a more extended observations campaign with more observation runs at good weather might provide valuable insights on AGN variability in the -- as yet poorly probed -- time regime from hours to days.

In principle, binning of light curves in the frame of a statistical multi-scale analysis can provide additional information and enhance the chance of detecting variability, especially weak flares that span many scans in time but appear insignificant when studying the original light curve \citep{KimJY2013}. However, in this case one has to sacrifice the time resolution which is critical in probing the shortest time scales of variability. As suggested by the periodogram analysis, all our lightcurves are consistent with being white noise signals, meaning that the amplitude of variations is the same on all time scales -- meaning that a binning of our light curves would not provide additional information.

\section{Summary and Conclusions}

We performed photometric radio single dish observations of four AGN -- 3C~111, 3C~279, 3C~454.3, and BL Lac -- using the three KVN 21-meter antennas at four different frequencies (22, 43, 86, and 129 GHz) in order to probe AGN intra-day variability. Observations were done in an ``anti-correlated'' pointing mode -- with always at least one antenna of a pair of antennas pointing at the target -- to construct long and undisrupted light curves which are of high importance in examining short time scale variability. In general, we achieved a time resolution of about one minute.

We were able to derive high-quality light curves for 3C~111, 3C~454.3, and BL Lacertae at 22 and 43 GHz, and for 3C~279 at 86 GHz, between May 2012 and April 2013. We found upper limits on flux variability ranging from $\sim$1.6\% to $\sim$7.6\%. The upper limits on the derived brightness temperatures exceed the inverse Compton limit by three to six orders of magnitude. From our results, plus comparison with data obtained by the UMRAO, we conclude that we have not detected source-intrinsic variability which would have to occur at sub-per cent levels.


\acknowledgments

We are grateful to all staff members of KVN who helped to operate the telescopes. The KVN is a facility operated by the Korea Astronomy and Space Science Institute. The KVN operations are supported by KREONET (Korea Research Environment Open NETwork) which is managed and operated by KISTI (Korea Institute of Science and Technology Information).
This project has been supported by the Korean National Research Foundation (NRF) via Basic Research Grant 2012-R1A1A-2041387.
This work made use of the NASA/IPAC Extragalactic Database (NED).


\end{document}